%% file: ANA-EXOT-2019-39-PAPER.tex
\documentclass[cernpreprint, atlasdraft=false, texlive=2023, UKenglish, orcidlogo, texmf, orcidlogo]{atlasdoc}
 
\usepackage{atlaspackage}
\usepackage{atlasbiblatex}
 
\usepackage{atlasphysics}
 
\ifthenelse{\boolean{AtlasDraft}}{\usepackage[output=true, shift=true]}{}
 
\graphicspath{{logos/}{figures/}}
 
\usepackage{comment}
 
\usepackage{ANA-EXOT-2019-39-PAPER-defs}

\AtlasTitle{Search for heavy Majorana or Dirac neutrinos and right-handed $W$ gauge bosons in final states with charged leptons and jets in $pp$ collisions at $\sqrt{s}=13$ TeV with the ATLAS detector}

\AtlasAbstract{
A search for heavy right-handed Majorana or Dirac neutrinos $N_{\mathrm{R}}$ and heavy right-handed gauge bosons $W_{\mathrm{R}}$ is performed in events with energetic electrons or muons, with the same or opposite electric charge, and energetic jets. The search is carried out separately for topologies of clearly separated final-state products (``resolved'' channel) and topologies with boosted final states with hadronic and/or leptonic products partially overlapping and reconstructed as a large-radius jet (``boosted'' channel). The events are selected from $pp$ collision data at the LHC with an integrated luminosity of 139 fb$^{-1}$ collected by the ATLAS detector at $\sqrt{s}$ = 13 TeV. No significant deviations from the Standard Model predictions are observed.
The results are interpreted within the theoretical framework of a left-right symmetric model, and lower limits are set on masses in the heavy right-handed $W_{\mathrm{R}}$ boson and $N_{\mathrm{R}}$ plane.
The excluded region extends to about $m(W_{\mathrm{R}}) = 6.4$~TeV for both Majorana and Dirac $N_{\mathrm{R}}$ neutrinos at $m(N_{\mathrm{R}})<1$~TeV.
$N_{\mathrm{R}}$ with masses of less than 3.5 (3.6)~TeV are excluded in the electron (muon) channel at $m(W_{\mathrm{R}})=4.8$~TeV for the Majorana neutrinos,
and limits of $m(N_{\mathrm{R}})$ up to  3.6~TeV for $m(W_{\mathrm{R}}) = 5.2$ (5.0)~TeV in the electron (muon) channel are set for the Dirac neutrinos. These constitute the most stringent exclusion limits to date for the model considered.
}
 
\author{The ATLAS Collaboration}
 
\AtlasRefCode{EXOT-2019-39}
 
\PreprintIdNumber{CERN-EP-2023-034}

\arXivId{2304.09553}
 
\HepDataRecord{ins2652625}
 
\AtlasJournal{EPJC}
\AtlasJournalRef{Eur. Phys. J. C 83 (2023) 1164}
\AtlasDOI{10.1140/epjc/s10052-023-12021-9}

\hypersetup{pdftitle={ATLAS document},pdfauthor={The ATLAS Collaboration}}
 
\begin{document}
 
\maketitle

\section{Introduction}

The very small mass of neutrinos is one of the biggest puzzles in the Standard Model (SM)
of particle physics. The seesaw mechanism~\cite{Minkowski:1977sc, Yanagida:1979as, GellMann:1980vs} is a proposed solution,
in which the light neutrinos acquire their Majorana masses through heavy right-handed neutrinos. From the effective field theory point of view, this is equivalent to dimension-5 operators~\cite{Weinberg:1979sa} through electroweak
symmetry breaking. Several types of the seesaw mechanism are proposed; for example,
Type-I with right-handed neutrinos~\cite{Minkowski:1977sc, Yanagida:1979as, GellMann:1980vs},
Type-II with a scalar triplet~\cite{Magg:1980ut,PhysRevD.22.2860,PhysRevD.23.165} and Type-III with at least two fermion triplets~\cite{Magg:1980ut, Foot:1988aq}
scenarios. Type-I and Type-II models can further be embedded into the Left-Right Symmetric Model
(LRSM)~\cite{Pati:1974yy, Mohapatra:1974gc, Senjanovic:1975rk}. The LRSM attempts to explain the broken parity symmetry of the weak interaction in the SM and can introduce, depending on the form of the model, right-handed counterparts to the $W$ and $Z$ bosons (\WR\ and $Z_R$), and right-handed heavy
neutrinos \NR\ as the parity gauge partners of the corresponding left-handed neutrino fields.
 
\begin{figure}[h]
\centering
\includegraphics[width=0.8\textwidth, keepaspectratio]{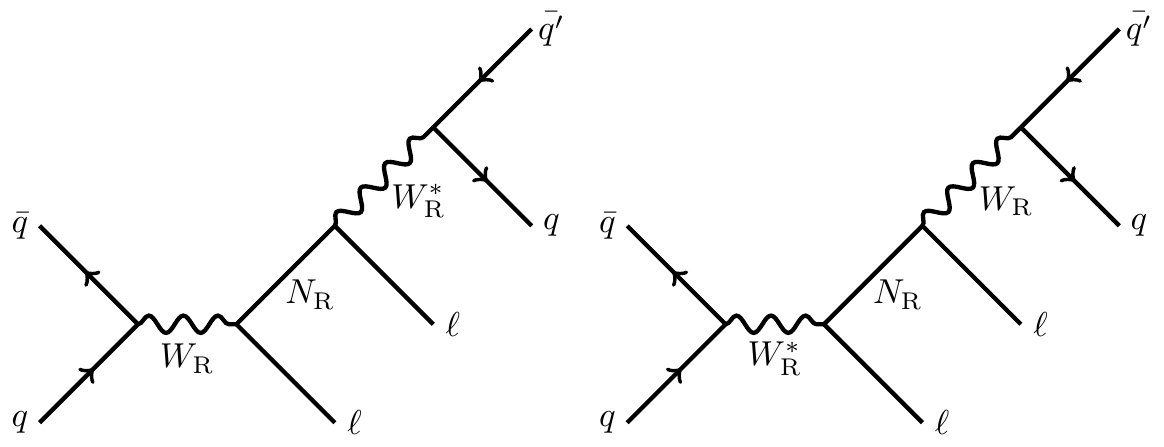}
\caption{
The Keung-Senjanovi\'c process for the m(\WR) > m(\NR) (left) and
the m(\NR) > m(\WR) (right) cases. The asterisk ($^*$) on the \WR\ denotes an off-shell particle.
In the case of Majorana \NR, the final state appears with two same-sign charged leptons 50\% of the time, violating the lepton number conservation.
}
\label{fig:WRfeyn}
\end{figure}

In this analysis, a search for \WR\ bosons decaying to \NR\ and a charged lepton $\ell^\pm$ in proton-proton ($pp$) collisions at a centre-of-mass energy $\sqrt{s}=13~\TeV$ with the ATLAS detector is presented,
where $\ell^\pm$ denotes an electron/positron ($e^{\pm}$)
or a muon ($\mu^{\pm}$). The exact process of interest is the Keung-Senjanovi\'c (KS) process \cite{Senjanovic:1975rk}.
In the case where the mass of \WR\ is
larger than the \NR\ mass, $\mWR > \mNR$, \NR\ decays into a charged lepton and an off-shell \WR ($\WR^{*}$). 
In the case of $\mWR < \mNR$, a  \WR\ produced off-shell decays into a pair of \NR\ and $\ell$, and the \NR\ subsequently
decays via the \WR\ resonant state. The charge of the final state leptons is a key feature of the analysis, as it determines the Dirac or Majorana nature of \NR.
The focus of the search is on hadronic decays of the final-state $\WR^{(*)}$ because of their high branching fractions.
The leading-order Feynman diagrams for the KS process targeted by this analysis are shown in Figure~\ref{fig:WRfeyn}.
Depending on the target mass range, two types of analyses are
performed; one requiring that the two quarks in the final state are clearly separated geometrically and are reconstructed as two separate jets (hereafter labelled as
the ``resolved'' channel), with the second one targeting the $\mWR \gg \mNR$ region where
particles from the \NR\ decay are merged due to the Lorentz boost, and are reconstructed as a single large-radius (large-$R$) jet  (the ``boosted'' channel)~\cite{Mitra_2016,PhysRevD.62.013001}.
 
Multiple experimental results can give constraints on the target signal of this analysis~\cite{Lindner:2016lxq, Biswal:2017nfl}.
The LRSM can enhance some low-energy processes, such as
$K$-$\overline{K}$ and $B_{d,s}$-$\overline{{B}_{d,s}}$ oscillations, via \WR\ contributions in the box diagram.
The lower limit on \mWR\ from the latest mixing results is about 3~\TeV~\cite{Bertolini:2014sua,Maiezza:2014ala}, assuming equal
mixing matrices for left- and right-handed quarks.
A Majorana right-handed electron neutrino and \WR\ boson can contribute to the neutrinoless
double beta decay ($0\nu \beta\beta$) diagram. Assuming purely right-handed contributions to the $0\nu\beta\beta$
decay, the non-observation of this lepton number violating process can be used to set limits on the
\WR\ and \NR\ masses in the LRSM. As an example, at $\mWR = 3~\TeV$ (5~TeV), the upper limit on the \NR\ mass is about
180~\GeV\ (20~\GeV)~\cite{GERDA, KamLAND-Zen}.
 
The $pp$ collisions at the Large Hadron Collider (LHC) offer a window into a unique phase space for this search, allowing the exploration of \mNR\ in a range from $O(100)$~\GeV\ to a few TeV.
Searches by the ATLAS~\cite{EXOT-2012-24,EXOT-2017-18,EXOT-2018-15} and
CMS~\cite{CMS-EXO-11-091,CMS-EXO-17-011,CMS-EXO-17-016,CMS-EXO-17-028,CMS-EXO-20-002}
collaborations have excluded signals in the LRSM with \mWR\ up to about 4.7~\TeV\ and 5~\TeV\ in the electron and muon final states, respectively, for a \mNR\ range from 100~\GeV\ to 3~\TeV.


\section{Signal model}
\label{sec:signal_model}

The theoretical framework of LRSM
offers the prediction of a right-handed charged current and a mass-generating mechanism for light and heavy neutrinos.
The small left-handed neutrino masses are naturally explained via right-handed neutrinos in the Type-I seesaw mechanism, or  via SU(2)-triplet scalars in
the Type-II seesaw model. Both Type-I and Type-II contributions can coexist in the LRSM.
 
In the minimal LRSM, the left-handed (i.e. SM-like) neutrinos as well as the right-handed
neutrinos are predicted to be Majorana particles.
The model thus features the violation of the global lepton number symmetry.
In the target KS process, same- and opposite-sign lepton pairs would be equally observed in a 50\%-50\% admixture of signal events.
In the LRSM variants that include the inverse seesaw mechanism~\cite{Mohapatra:1986aw, Mohapatra:1986bd, Chen:2011hc, BhupalDev:2015kgw}, the \NR\ neutrinos are pseudo-Dirac particles formed by two Majorana particles with identical masses~\cite{Wolfenstein:1981kw}.
For simple versions of LRSMs incorporating the inverse seesaw mechanism, lepton-number-violating processes are not expected~\cite{Das:2017hmg}.
In this paper we explore both ``Dirac'' and ``Majorana'' interpretations of the LRSM.


\section{ATLAS detector}

\newcommand{\AtlasCoordFootnote}{
ATLAS uses a right-handed coordinate system with its origin at the nominal interaction point (IP)
in the centre of the detector and the \(z\)-axis along the beam pipe.
The \(x\)-axis points from the IP to the centre of the LHC ring,
and the \(y\)-axis points upwards.
Cylindrical coordinates \((r,\phi)\) are used in the transverse plane,
\(\phi\) being the azimuthal angle around the \(z\)-axis.
The pseudorapidity is defined in terms of the polar angle \(\theta\) as \(\eta = -\ln \tan(\theta/2)\).
Angular distance is measured in units of \(\Delta R \equiv \sqrt{(\Delta\eta)^{2} + (\Delta\phi)^{2}}\).}

The ATLAS experiment~\cite{PERF-2007-01} at the LHC is a multipurpose particle detector
with a forward--backward symmetric cylindrical geometry and a near \(4\pi\) coverage in
solid angle.\footnote{\AtlasCoordFootnote}
It consists of an inner tracking detector surrounded by a thin superconducting solenoid
providing a \SI{2}{\tesla} axial magnetic field, electromagnetic and hadron calorimeters, and a muon spectrometer (MS).
The inner tracking detector covers the pseudorapidity range \(|\eta| < 2.5\).
It consists of silicon pixel, silicon microstrip, and transition radiation tracking detectors.
Lead/liquid-argon (LAr) sampling calorimeters provide electromagnetic (EM) energy measurements
with high granularity.
A steel/scintillator-tile hadron calorimeter covers the central pseudorapidity range (\(|\eta| < 1.7\)).
The endcap and forward regions are instrumented with LAr calorimeters
for both the EM and hadronic energy measurements up to \(|\eta| = 4.9\).
The MS surrounds the calorimeters and is based on
three large superconducting air-core toroidal magnets with eight coils each.
The field integral of the toroids ranges between \num{2.0} and \SI{6.0}{\tesla\metre}
across most of the detector.
The MS includes a system of precision tracking chambers and fast detectors for triggering.
A two-level trigger system is used to select events.
The first-level trigger (L1) is implemented in hardware and uses a subset of the detector information
to accept events at a rate below \SI{100}{\kHz}.
This is followed by a software-based high-level trigger (HLT) that
reduces the accepted event rate to \SI{1}{\kHz} on average
depending on the data-taking conditions.
An extensive software suite~\cite{ATL-SOFT-PUB-2021-001} is used in the reconstruction and analysis of real
and simulated data, in detector operations, and in the trigger and data acquisition systems of the experiment.


\section{Dataset and simulated event samples}

The data used in this analysis
were collected with the ATLAS detector between 2015 and 2018, and correspond to an integrated luminosity of 139~\ifb.
The average number of $pp$ interactions per bunch crossing (``pile-up'') in the dataset is 33.7.
Only high-quality data, collected when the LHC has declared the beams to be stable and all of the ATLAS subdetectors are reported to be operating well, are analysed ~\cite{DAPR-2018-01, ATLAS-CONF-2015-029}.
All events are required to have a vertex with at least two associated ID tracks with $\pt>500~\MeV$~\cite{ATL-PHYS-PUB-2015-026, PERF-2015-01}.
The one with the highest $\sum \pt^{2}$ of the associated tracks is selected as the primary vertex.

The dataset was collected by single or dilepton triggers with a variety of transverse momentum (\pt) and isolation requirements, which depend on the data-taking period.
In both resolved and boosted channels, only events in the region of trigger efficiency plateau~\cite{TRIG-2018-01,TRIG-2018-05} are used.
Trigger thresholds are no higher than (24 GeV, 24 GeV) for dielectrons and (22 GeV, 8 GeV) for dimuons in the resolved channel, and 140 GeV for electrons and 50 GeV for muons in the boosted channel.
The trigger efficiency for most signal points for both resolved and boosted channels ranges from 90 to 95\%, dropping to 85\% in the $m(\WR) \sim m(\NR)$ region where the compressed phase space affects the charged lepton kinematics.
For the resolved analysis, the $e\mu$ trigger is used for the \ttbar\ background estimation.
In addition, for the study of mis-identified leptons in the resolved analysis (Section~\ref{sec:resolved}) some non-isolated, heavily-prescaled single lepton triggers were also used.
 
Monte-Carlo (MC) simulated events are used to optimise the event selections, and to estimate the background contribution and the systematic uncertainties.
For all MC events, the response of the ATLAS detector is simulated using the \GEANT toolkit~\cite{agostinelli2003geant4}.
The same reconstruction and trigger algorithms are applied for data and simulated events, using the default ATLAS software~\cite{SOFT-2010-01}.
Multiple overlaid $pp$ collisions are simulated with the soft QCD processes of
\texttt{Pythia}~8.230~\cite{Sjostrand:2014zea} using the A3 set of tuned parameters~\cite{ATL-PHYS-PUB-2016-017} and the NNPDF2.3LO parton distribution function (PDF) set~\cite{Ball:2012cx}.
MC events are reweighted so that the distribution of the average number of interactions per bunch crossing agrees with the data (``pile-up reweighting'').
 
Table~\ref{tab:background_mc_samples} summarises the simulation packages used for the SM background processes.
More details are found in other recent ATLAS publications, for example Refs.~\cite{STDM-2019-22,Aad:2798348,ATL-PHYS-PUB-2017-007,ATL-PHYS-PUB-2018-005}.
The multijet and $\gamma$+jets samples are used only in the boosted channel.
 
The signal events for $pp \to \WR^{(*)} \to \ell\ell qq'$ in the LRSM are generated at leading-order (LO) using FeynRules~\cite{alloul2014feynrules} implemented in \texttt{MadGraph5\_aMC@NLO}~\cite{alwall2014automated} and further modified as described in Ref.~\cite{Nemevsek:2018bbt}, where $\ell$ includes the tau lepton.
The generated events are interfaced with \texttt{Pythia}~8.230~\cite{Sjostrand:2014zea} for parton showering and hadronisation.
The A14 parameter set is used for tuning the shower~\cite{ATL-PHYS-PUB-2014-021}.
The NNPDF3.1NLO~\cite{Ball:2014uwa} PDF set enters in the matrix element
calculation and the NNPDF2.3LO is used in the parton shower.
 
Lepton flavour mixing, albeit possible in principle in the LRSM, is not considered in this analysis.
The branching fractions for the electron, muon and tau channels for \WR\ decays to \NR\ of equal masses are assumed to be exactly one-third for each flavour due to lepton universality.
For sufficiently heavy \WR\ masses, $\mWR > m_{t}$, the hadronic final state also includes the $\WR\ \to tb$ channel.
However, $b$-jets are vetoed in this search to reduce the top background contamination, effectively limiting the phase space to hadronic $\WR$ decays excluding $b$-quarks.
At very low \NR\ masses of less than 50~\GeV, the decay length of \NR\ is greater than 1~mm~\cite{Nemevsek:2018bbt}. This mass range can be explored by dedicated ATLAS and CMS analyses requiring a displaced vertex, and is beyond the scope of this paper. The \mWR\ and \mNR\ are sampled from 1~\TeV\ to 7~\TeV\ and from 50~\GeV\ to 4~\TeV, respectively, at intervals of about 500~\GeV.
In the $\mWR > 10 \times \mNR$ region, which is the focus of the boosted channel, the \NR\ mass is further sampled in steps of 100~GeV.
The simulation assumes a Majorana \NR, giving a 50\% mixture of same-sign and opposite-sign lepton pairs. For the Dirac neutrino case, only the opposite-sign events are used in the analysis as no other differences are expected, and the production cross section is adjusted appropriately.
 
\begin{table}[tbhp]
\caption{Summary of the simulated background events used in this analysis, including name of the event generator, accuracy in QCD when the matrix element is calculated, parton shower algorithm, underlying event (UE) tune, and accuracy in QCD when the cross section is calculated. NLO, NNLO and NNLL denote the next-to-leading order, next-to-next-to-leading order, and next-to-next-to-leading logarithm, respectively,
and $n_{\textrm{parton}}$ is the number of partons in the matrix-element calculation. $V$ denotes a $W$ or $Z$ boson.}
\label{tab:background_mc_samples}
\begin{center}
\resizebox{\textwidth}{!}{
\begin{tabular}{llllll}
\toprule
Sample & Generator &  Matrix Element &  Parton shower &  UE tune & cross section \\
\midrule
$Z\to \ell\ell$ + jets and	& \texttt{Sherpa} 2.2.11~\cite{Bothmann:2019yzt}	& NLO@$n_{\textrm{parton}}\leq 2$	& \texttt{Sherpa} default~\cite{Schumann:2007mg}	&  \texttt{Sherpa} default 	& NNLO~\cite{Anastasiou:2003ds} \\
$W\to \ell\nu$ + jets	&						& LO@$n_{\textrm{parton}}=3$,4,5	&				&					& 	 \\
\midrule
\ttbar	\ and single-$t$			& \texttt{Powheg-Box} v2~\cite{Frixione:2007nw, Nason:2004rx, Frixione:2007vw, Alioli:2010xd}	& NLO	& \texttt{Pythia} 8.230 &  A14 tune~\cite{ATL-PHYS-PUB-2014-021} & NNLO+NNLL~\cite{Beneke:2011mq,Cacciari:2011hy,Baernreuther:2012ws,Czakon:2012zr,Czakon:2012pz,Czakon:2013goa,Czakon:2011xx} \\
\midrule
$VV$ and $VVV$
& \texttt{Sherpa} 2.2.1 or 2.2.2~\cite{Bothmann:2019yzt}	& NLO@$n_{\textrm{parton}}\leq 1$	& \texttt{Sherpa} default & \texttt{Sherpa} default 	& NLO ($VV$)~\cite{Chiesa_2020} / \\
&						& LO@$n_{\textrm{parton}}=2$,3	& 				&					&NNLO ($VVV$)~\cite{KALLWEIT2021136013} 	 \\
\midrule
$\gamma$ + jets & \texttt{Sherpa} 2.1~\cite{Bothmann:2019yzt} & LO@$n_{\textrm{parton}} \leq 3$ & \texttt{Sherpa} default & \texttt{Sherpa} default  &  LO \\
\midrule
Multijet & \texttt{Pythia} 8.230~\cite{Sjostrand:2014zea}	& LO for dijet events & \texttt{Pythia} 8.230   & A14 tune~\cite{ATL-PHYS-PUB-2014-021}	& LO \\
\bottomrule
\end{tabular}
}
\end{center}
\end{table}


\section{Object reconstruction}
\label{sec:object_definition}

The definitions of the physics objects used in this analysis are summarised in Table~\ref{tab:object_selection}.
Physics objects in the events (electrons, muons, jets, and missing transverse momentum) are separately defined for the resolved and boosted analyses.
The orthogonality between the two channels is not enforced: the same event can be used in both analyses, which are not statistically combined.
\begin{table}[tbh]
\caption{Definitions of the electrons, muons, small-$R$ and large-$R$ jets,
used in this analysis. Optimisations of the object selections are performed separately for the resolved and boosted analyses. See text about the definitions of
`Baseline' and `Leading' leptons as well as leptons for `Fake estimation'.}
\label{tab:object_selection}
\centering
\scalebox{0.7}{
\begin{tabular}{lccc|ccc}
\hline
& & \multicolumn{2}{c|}{\textbf{Resolved}} & \multicolumn{3}{c}{\textbf{Boosted}} \\
\hline
& & \textbf{Baseline} & \textbf{Fake estimation} & \textbf{Baseline} & \textbf{Leading} & \textbf{Fake estimation} \\
\cline{2-7}
& $\left| \eta \right|$    & \multicolumn{5}{c}{$(0, 1.37]$ or $[1.52,2.47]$} \\
& \pt (GeV)       & \multicolumn{2}{c|}{$>25$} & $>25$ & \multicolumn{2}{c}{$>200$}\\
\textbf{Electrons}            & Quality    & \texttt{Tight}  & \texttt{Loose}     & \texttt{Medium} & \multicolumn{2}{c}{\texttt{Tight}} \\
& Isolation & Loose & Fail \texttt{Loose} or \texttt{Tight} & \texttt{Loose} &  \texttt{HighPtCaloOnly} & \texttt{Loose} but fail \\
&               &            &                                &            &                             & \texttt{HighPtCaloOnly} \\
\hline
& & \textbf{Baseline} & \textbf{Fake estimation} & \textbf{Baseline} &  \textbf{Leading} & \\
\cline{2-7}
& \pt (GeV)       & \multicolumn{2}{c|}{$>25$} & $>28$ & $>200$ & ---\\
\textbf{Muons}              & $\left| \eta \right|$    & \multicolumn{2}{c|}{$<2.5$} & \multicolumn{2}{c}{$<2.5$} & ---\\
& Quality   & \multicolumn{2}{c|}{\texttt{High-\pt} if pT $> 300~\GeV$ else \texttt{Medium}} & \texttt{Medium}     & \texttt{Tight} & ---\\
& Isolation & \texttt{FixedCutTightTrackOnly} & fail \texttt{FixedCutTightTrackOnly} &  --- & \texttt{Tight}  & ---\\
\hline
 
& \pt (GeV)       & \multicolumn{5}{c}{$>20$} \\
\raisebox{0.5 em}{\textbf{Small-$R$ jet}}            & $\left|\eta\right|$   & \multicolumn{5}{c}{$<2.5$} \\
\hline
 
& \pt  (GeV)      & \multicolumn{2}{c|}{---} & \multicolumn{3}{c}{$>200$} \\
\raisebox{0.5 em}{\textbf{Large-$R$ jet}}            & $\left|\eta\right|$    & \multicolumn{2}{c|}{---} & \multicolumn{3}{c}{$<2$} \\
\hline
\end{tabular}
}
\end{table}

Electrons are reconstructed as ID tracks, matched to energy clusters in the EM calorimeter
within $\left|\eta\right| < 2.47$~\cite{EGAM-2018-01}.
The electron candidates in the crack region of the EM calorimeter ($1.37 < \left|\eta\right| < 1.52$) are discarded.
The electron identification utilises a multivariate likelihood-based discriminant that exploits the shower shapes in the EM calorimeter and the
associated track properties. There are `\texttt{Loose}', `\texttt{Medium}' and `\texttt{Tight}' identification working points described in Ref.~\cite{EGAM-2018-01}.
To further suppress the mis-identified electrons, an isolation criterion can be applied to the electron candidates.
Several isolation working points are defined in Ref.~\cite{EGAM-2018-01} e.g. `\texttt{Loose}' and `\texttt{HighPtCaloOnly}'.
Slightly different \pt\ thresholds are used across the resolved and boosted channels to ensure the operation at a region of constant trigger efficiency.
For the resolved analysis, electrons are required to have $\pt>25~\GeV$ and to satisfy \texttt{Tight} identification and \texttt{Loose} isolation criteria.
For the boosted analysis, electrons must have $\pt>25~\GeV$, and satisfy \texttt{Medium} identification and \texttt{Loose} isolation criteria.
Additional cuts are applied for the highest-\pt\ electron candidate in the boosted channel (Section~\ref{subsec:event_selection:boosted}), to ensure that $\pt>200~\GeV$, and to satisfy the \texttt{Tight} identification and \texttt{HighPtCaloOnly} isolation criteria.
Tighter isolation criteria are required in the boosted channel because the multijet process with a mis-identified electron is a major source of background.
 
Muons are reconstructed from MS tracks matching ID tracks in the $\left| \eta \right| < 2.5$ region.
There are several muon identification working points described in Ref.~\cite{PERF-2015-10}, namely `\texttt{Loose}', `\texttt{Medium}' ,`\texttt{Tight}' and `\texttt{High-\pt}'.
The resolved channel uses muons with $\pt>25~\GeV$ satisfying the \texttt{Medium} identification working point.
For muons with $\pt>300~\GeV$, the \texttt{High-\pt} working point requirements must be met.
For the boosted channel, muons with $\pt>28~\GeV$ satisfying the \texttt{Medium} working point are used.
As with electrons, several muon isolation working points are defined in Ref.~\cite{PERF-2015-10},
e.g. `\texttt{FixedCutTightTrackOnly}' and `\texttt{Tight}'.
The resolved channel requires muons to satisfy the \texttt{FixedCutTightTrackOnly} working point.
In the boosted channel, no isolation criterion is required on the muon from the $\NR \to \mu qq'$ decay,
as it is not expected to be clearly isolated from the hadrons.
Further selection cuts are applied for the highest-\pt\ muon candidate in the event, namely
to have $\pt>200~\GeV$ and satisfy the \texttt{Tight} identification and \texttt{Tight} isolation criteria (Section~\ref{subsec:event_selection:boosted}).
 
For both electrons and muons, track-to-vertex association requirements are employed by using the impact parameter observables.
The longitudinal impact parameter of the lepton track, $z_{0}$, is required to satisfy $\left| z_{0} \sin\theta \right| < 0.5$~mm,
where $\theta$ is the polar angle of the track. In addition, the transverse impact parameter divided by its uncertainty, $|d_{0}|/\sigma(d_{0})$, is
required to be less than 5 (3) for electrons (muons).
The leptons' reconstruction, identification, isolation and trigger efficiencies differ slightly between simulation and data. The simulation is corrected with scale factors to match the data efficiencies~\cite{EGAM-2018-01, PERF-2015-10}.
 
Apart from the baseline leptons described above, leptons with modified requirements are employed for control samples for the estimation of jets misidentified as leptons or non-prompt leptons from decays of hadrons.
In the resolved channel, the isolation requirement is inverted and the \texttt{Loose} identification is employed.
In the boosted channel, the fraction of events containing fake muons surviving the $\pt>200~\GeV$ requirement for the leading lepton is negligibly small.
To estimate the hadrons mis-identified as electrons, electron candidates satisfying the \texttt{Loose} but failing the \texttt{HighPtCaloOnly} isolation criterion are used.
 
Jets are reconstructed from particle-flow objects~\cite{PERF-2015-09} using the anti-$k_{t}$ algorithm~\cite{Cacciari:2008gp, Fastjet} with radius parameter $R=0.4$.
They are referred to as `small-$R$ jets' hereafter.
A detailed description of the calibration of these jets is found in Ref.~\cite{JETM-2018-05}.
Only small-$R$ jets with $\pt>20~\GeV$ and $\left| \eta \right| < 2.5$ are considered.
small-$R$ jets with $\pt < 60~\GeV$ and $\left|\eta\right| < 2.4$ from pile-up interactions are suppressed using a jet-vertex tagging (JVT) discriminant~\cite{PERF-2014-03}.
small-$R$ jets containing $b$-flavoured hadrons ($b$-jets) are identified with the \texttt{DL1r} multivariate $b$-tagging algorithms~\cite{FTAG-2018-01, ATL-PHYS-PUB-2017-013}.
A working point to achieve 77\% efficiency for $b$-jets is employed, which has a rejection factor of 5 for $c$-jets and 169 for light jets. It is only used to veto events containing jets identified as $b$-jets.
 
The missing transverse momentum, with magnitude \met, is calculated using the baseline electrons and the small-$R$ jets in each event as described in Refs.~\cite{PERF-2016-07, ATLAS-CONF-2018-023} using the \texttt{Tight} working point.
The particle-flow track-based soft term, built from tracks that are matched to the primary vertex but not associated with any other objects, is also used in the calculation.
The \met\ is used only in the electron channel, and muons are not considered in the calculation, since the muon momentum resolution is significantly degraded in the high-\pt\ region and affects the \met\ calculation.
 
To avoid cases where the detector response to a single physical object is reconstructed as two different
final-state objects, e.g. an electron reconstructed as both an electron and a jet,
several steps are followed, as summarised in Table~\ref{tab:object_selection:or}. To improve the efficiency for muons from \NR\ decays
in the boosted analysis, the last step of muon-jet overlap removal is considered only for the highest-\pt\ muon in the event.
 
\begin{table}[ht]
\caption{
The order of overlap removal used in this analysis. $\Delta R$ is defined as $\Delta R = \sqrt{ \Delta y^{2} + \Delta \phi^{2}}$,
where $\Delta y$ and $\Delta \phi$ are the rapidity and azimuthal-angle differences between two objects.
The variables $\pt(e)$, $\pt(\mu)$, $\pt(j)$ and $\pt(\textrm{trk})$ are the transverse momenta of the corresponding electron,
muon, small-$R$ jets, and all tracks associated with the small-$R$ jet. The baseline definitions of electrons and muons are used. In the boosted channel, the last step of muon-jet overlap removal is considered only for the highest-$\pt$ muon in the event. Large-$R$ jets are reclustered from small-$R$ jets after the overlap removal.
}
\label{tab:object_selection:or}
\centering
\scalebox{0.9}{
\begin{tabular}{c l l l}
\toprule
Order & Object discarded & Object kept & \multicolumn{1}{c}{Matching condition} \\
\midrule
1. & Electron    & Electron & If two electrons share a track, discard the softer electron \\
2. & Muon        & Electron & If they share a track and the muon type is  \\
&		&		&	calorimeter-tagged ~\cite{PERF-2015-10} \\
3. & Electron    & Muon     & If they share a track with the remaining muon \\
4. & Small-$R$ jet  & Electron & $\Delta R < 0.2$, but step is skipped if jet is $b$-tagged \\
&                        &                & and $\pt(e) < 100~\GeV$ \\
5. & Electron    & Small-$R$ jet & $\Delta R < 0.4$ \\
6. & Small-$R$ jet  & Muon     & $\Delta R < 0.2$ , number of tracks associated to the jet $<$ 3,  \\
&		&		&	$\pt(\mu)/\pt(j) > 0.5$ and $\pt(\mu)/\sum \pt(\textrm{trk}) > 0.7$ \\
7. & Muon        & Small-$R$ jet & $\Delta R < 0.4$ \\
\bottomrule
\end{tabular}
}
\end{table}

In the boosted analysis, the hadronic jets from the \NR\ decay are reconstructed as a single large-$R$ jet.
The large-$R$ jet is reconstructed from the small-$R$ jets described above, using the anti-$k_t$ algorithm with $R=1.0$.
For such large-$R$ jets, `re-clustered' from small-$R$ jets,
the small-$R$ jet calibration can be propagated. Large-$R$ jets with $\pt > 200~\GeV$ and $\left|\eta\right| < 2.0$ are considered.


\section{Resolved channel: event selection and background estimate}
\label{sec:resolved}

\subsection{Event reconstruction and selection}

The resolved channel targets signals with a mass splitting between \WR\ and \NR, i.e. $\Delta m \equiv \mWR - \mNR$, up to $4~\TeV$.
Events passing the following requirements are filtered into datasets comprising signal regions (\texttt{SR}s),
namely regions where the presence of signal is hypothesised by the theoretical models and the analysis is expected to have high sensitivity.
The definitions of the \texttt{SR}s are summarised in Table~\ref{tab:event_selection:resolved:definition}.
Events are required to have exactly two baseline, same-flavour charged leptons, with $\pt > 40~\GeV$ for the leading lepton, and at least two small-$R$ jets with $\pt > 100~\GeV$.
To suppress the $Z$+jets background, the dilepton invariant mass must satisfy $m_{\ell\ell} > 400~\GeV$.
Since the focus of this search is the high \WR\ mass region,
the following requirements are imposed on the dijet invariant mass ($m_{jj}$), and the scalar sum of the transverse momentum of the leptons and the two leading small-$R$ jets ($h_{\textrm{T}}$): $m_{jj}>110~\GeV$ and $h_{\textrm{T}}>400~\GeV$.
Events are further separated into two categories based on the charge product of the two leptons: same- (\rSRSS) or opposite-sign (\rSROS) lepton pairs, where \texttt{r} denotes the resolved channel and \texttt{SR} the classification as signal region.
The relatively small SM background in \rSRSS\ increases the search sensitivity for Majorana \NR\ signals where half of the events are expected to appear with same-sign lepton pairs.
The $\Delta R$ between the two leptons is required to be less than 3.9 in \rSRSS\ to reduce some mismodelling of the simulated diboson background without a signal efficiency loss.
Finally, the \texttt{SR}s are separated into electron and muon channels.
The four resulting \texttt{SR}s (\rSROSTwoe, \rSROSTwomu, \rSRSSTwoe\ and \rSRSSTwomu) are orthogonal to each other and combined for a statistical analysis in the Majorana \NR\ interpretation.
For the Dirac \NR\ interpretation, only \rSROSTwoe\ and \rSROSTwomu\ are considered.
 
\begin{table}[htb]
\centering
\caption{Definition of signal regions in the resolved channel. The baseline definitions of electrons and muons are used.}
\label{tab:event_selection:resolved:definition}
\begin{tabular}{lcccc}
\hline
Variable                 & \rSRSSTwoe & \rSRSSTwomu & \rSROSTwoe & \rSROSTwomu \\ \hline
Number of electrons      & 2    &   0       &   2	&	0  \\ \hline
Number of muons          & 0    &   2       &   0	&	2  \\ \hline
Lepton charge          & \multicolumn{2}{c}{same sign} & \multicolumn{2}{c}{opposite sign} \\
\hline
Leading lepton $\pt$ [\GeV]    & \multicolumn{4}{c}{$>40$} \\
\hline
Dilepton mass $m_{\ell\ell}$ [\GeV]          &  \multicolumn{4}{c}{$>400$} \\
\hline
$\Delta R_{\ell\ell}$      & \multicolumn{2}{c}{$<3.9$} & \multicolumn{2}{c}{---} \\
\hline
Number of small-$R$ jets with $\pt>100~\GeV$ & \multicolumn{4}{c}{$\geq 2$} \\
\hline
Number of $b$-tagged jets  & \multicolumn{4}{c}{0} \\
\hline
Dijet mass $m_{jj}$ [\GeV]           &  \multicolumn{4}{c}{$>110$} \\
\hline
$h_{\textrm{T}} \equiv \pt(\ell_1) + \pt(\ell_2) + \pt(j_1) + \pt(j_2)$ [\GeV]                    & \multicolumn{4}{c}{$>400$} \\
\hline
\end{tabular}
\end{table}
 
In \rSROS, the \WR\ mass is reconstructed by combining the two charged leptons and the two leading small-$R$ jets ($m_{\ell\ell jj}$), and used as the final discriminant for signal points in the $\mWR > \mNR$ region.
The $0 < m_{\ell\ell jj} < 5~\TeV$ region is scanned with a step size of 500~\GeV, with one additional overflow bin that includes all events with $m_{\ell\ell jj}>5~\TeV$.
For signals with $\mWR < \mNR$, the $m_{jj}$ is used instead as a discriminant, with a similar 500~\GeV\ binning in $0 < m_{jj} < 3~\TeV$, and one additional overflow bin for $m_{jj}>3~\TeV$. The significance of yields is checked in every bin for both discriminants ($m_{\ell\ell jj}$ and $m_{jj}$) and the appropriate one is used for each ($\mWR$, $\mNR$) hypothesis for the interpretation.
In the smaller-background \rSRSS\ subset, the variable $h_{\textrm{T}}$ offers slightly better sensitivity and is used instead as the final discriminant. This variable is segmented into five bins: 400--600~\GeV, 600--1000~\GeV, 1--1.5~\TeV, 1.5--2.2~\TeV, and $> 2.2~\TeV$, which take into consideration the signal resolution and MC statistical uncertainty in each bin.


\subsection{Background estimation}

The composition of the SM background is substantially different between \rSROS\ and \rSRSS,
requiring different estimation techniques in the two categories. SM backgrounds
containing two prompt leptons are estimated using
simulations in both cases.
An overall normalisation correction obtained from the observed data in control regions (\texttt{CR}s) is used for the main background sources: $Z$+jets, diboson, and \ttbar\ processes.
\texttt{CR}s are kinematically similar to \texttt{SR}s, and enriched in the process for which the normalisation correction is applied. Signal contamination is very low (typically 1-2\%).
A different normalisation correction factor is assigned to each process, and typically treated as a free parameter in the fit discussed in Section~\ref{sec:results}.
The validity of the normalisation correction factor is confirmed in validation regions (\texttt{VR}s).
The component containing non-prompt or mis-identified leptons is estimated separately by employing a data-driven method.
The definitions of \texttt{SR}s, \texttt{CR}s and \texttt{VR}s for the resolved channel are visualised in Figure~\ref{fig:background_estimation:resolved:region_definition}.
 
\begin{figure}[htb]
\centering
\includegraphics[width=0.6\textwidth, keepaspectratio]{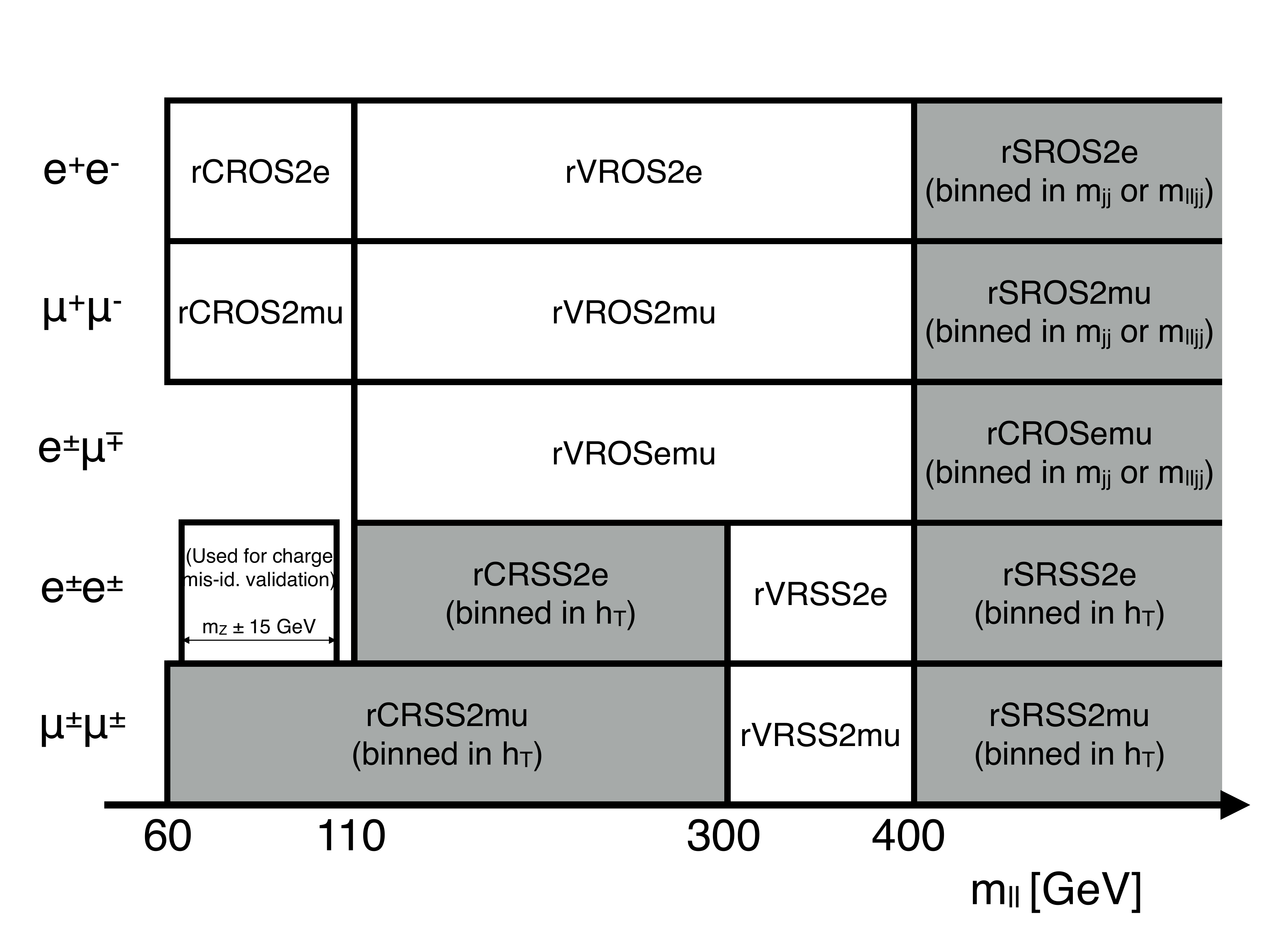}
\caption{
Resolved channel: Schematic view of \texttt{SR}/\texttt{CR}/\texttt{VR}s definitions.
The grey coloured regions are contributing simultaneously to the final fit in the electron or muons channels.
The remaining regions are used to verify the background estimations (see text for details).
}
\label{fig:background_estimation:resolved:region_definition}
\end{figure}

\subsubsection{Opposite-sign category}
\label{sec:ros_bgd}
The main SM backgrounds contributing to \rSROS\ are events with top quarks
(mainly \ttbar) and $Z$+jets production, with contributions of 41\% and 45\% respectively in
\rSROSTwoe\ and 41\% and 48\% in \rSROSTwomu.
Minor contributions arise from diboson processes (mainly $ZW \to \ell\ell jj$ and $ZZ \to \ell\ell jj$).
 
The normalisation factor for $Z$+jets is estimated in a control region defined by the same selection as \rSROS\ with a modification in the dilepton mass selection: $60 < m_{\ell\ell} < 110~\GeV$ (\rCROS).
There is a known issue in the jet multiplicity description of the $Z$+jets MC sample observed in a previous analysis~\cite{EXOT-2017-18},
leading to the mis-modelling of the dijet invariant mass $m_{jj}$.
Following the prescription described therein, a data-driven method is employed to correct for this effect. Electron and muon channels are combined and used to derive the $m_{jj}$-dependent correction factor.
The \rCROS\ data is used after subtracting the top-quark (3\%) and other minor prompt-lepton (2\%) background contributions evaluated with MC simulation, leaving a distribution with a size of about 94\% of the original dataset.
The normalisation of the $Z$+jets MC sample is then scaled to match the resulting \rCROS\ distribution.
The data-to-$Z$+jets MC ratio, $r$, reaches a maximum value of about 1.05 at $m_{jj}=300~\GeV$, and
decreases at the higher-mass region to reach a value of $\sim$0.7 at $m_{jj}=3~\TeV$.
The ratio $r$ is parameterised by the Novosibirsk function, as discussed in Ref.~\cite{EXOT-2017-18}.
The $r$ is used to apply an event-by-event correction to the $Z$+jets simulated sample in the \texttt{OS} channel.
The method corrects not only the $m_{jj}$ shape but also the normalisation of $Z$+jets in the simulation.
Therefore, \rCROS\ is not used in the final distribution fit discussed in Section~\ref{sec:results}.
 
The \ttbar\ background is estimated with the same selection as \rSROS, but by requiring a different flavour for the two charged leptons i.e. $m_{e\mu} > 400~\GeV$ (\rCROSemu).
The normalisation factor is treated as a free parameter in the final fit, but determined mainly in \rCROSemu\ due to its high purity in \ttbar ($\sim$ 80\%), with dibosons being the other main component ($\sim$ 19\%).
 
The validity of the background estimations described above is checked in the validation region \rVROS, defined in an intermediate $m_{\ell\ell}$ region between \texttt{SR} and \texttt{CR}, i.e. $110<m_{\ell\ell}<400~\GeV$.
An additional validation region is defined with the same selection as \rSROS\ but with an $e\mu$ pair instead of two same-flavour charged leptons (\rVROSemu). Validation results are presented in Sec.~\ref{results:resolved}.
 
\subsubsection{Same-sign category}
Diboson processes (mainly $WZ \to \ell\ell$ and same-sign $W^{\pm}W^{\pm} \to \ell^{\pm}\nu\ell^{\pm}\nu$) are the main background source in \rSRSS. In particular, about 83\%\ of the background in \rSRSSTwomu\ can be attributed to dibosons.
In addition, there are non-negligible contributions from $Z (\to ee)$+jets and top-quark events with charge mis-identification in \rSRSSTwoe.
The background contributions from diboson and charge mis-identified processes in \rSRSSTwoe\ are 29\% and 39\%.
 
The diboson background is estimated with the same selection as \rSRSS, but by requiring $110 < m_{ee} < 300~\GeV$ or $60 < m_{\mu\mu} < 300~\GeV$ (\rCRSS). To increase the \texttt{CR} dataset size, no selection is applied on $h_{\textrm{T}}$ or $m_{jj}$. The validity of the background estimation technique is confirmed in the $300<m_{\ell\ell}<400~\GeV$ validation region (\rVRSS), with validation results shown in Sec.~\ref{results:resolved}.
 
The charge-mis-identified background consists of events with two opposite-sign electrons, one of which has its charge sign mis-measured. The most frequent cause of electron charge flip is the bremsstrahlung effect.
The probability of an electron undergoing charge mis-identification is measured in the data ~\cite{PERF-2017-01} and compared to the MC prediction.
A correction factor is then derived for the charge flip probability
and applied to simulated events featuring same-sign electron pairs.
The method is validated in a same-sign dielectron region with $|m_{ee} - m_{Z}| < 15~\GeV$, which does not overlap with the \rCRSS\ region ($110 < m_{ee} < 300~\GeV$).
The data agrees well with the corrected prediction within uncertainties.
This result confirms the validity of the correction factor for the charge-flip probability in a limited $m_{ee}$ range, but does not cover a possible mis-modelling of the extended $m_{ee}$ distribution.
Unlike opposite-sign $Z$+jets, the normalisation of the same-sign $Z$+jet sample is therefore free to float and determined mainly in \rCRSSTwoe.
 
Another smaller background contribution comes from
mis-identified leptons originating from non-isolated, non-prompt electrons and muons produced by secondary decays of light- or heavy-flavour hadrons.
Another significant component of fake electrons arises from photon conversions, as well as from jets that are misreconstructed as electrons.
The contribution of events with mis-identified leptons is estimated by a data-driven technique, known as the ``fake-factor'' method.
The fake factor $F$ is defined as the ratio of mis-identified leptons satisfying the baseline lepton criteria (labelled as \texttt{Tight} in this section) to those satisfying the selection criteria for fake estimation (\texttt{Loose}).
The factors are estimated separately for electrons and muons.
The fake factors calculated in Ref.~\cite{EXOT-2018-33} are used in this analysis, as common object definitions are employed.
Fake factors are measured in data as a function of lepton \pt\ and $\eta$ in control samples that are orthogonal to the \texttt{SR}, \texttt{VR}, and \texttt{CR}s used in this analysis.
 
The contribution of mis-identified leptons in \texttt{SR}, \texttt{VR}, and \texttt{CR}s  is estimated as follows.
The calculated $F$ is applied to events satisfying the same selection criteria as in \texttt{SR}, \texttt{VR}, and \texttt{CR}s but by lowering the lepton ID requirement from \texttt{Tight} to \texttt{Loose} for at least one lepton.
The signal contamination in these regions is found to be small (typically 1-2\%) and ignored here.
Three categories are considered separately, depending on the identification criteria satisfied by the leading and sub-leading leptons: they are the $LT$ (\texttt{Loose} - \texttt{Tight}), $TL$ (\texttt{Tight} - \texttt{Loose}) and $LL$ (\texttt{Loose} -  \texttt{Loose}) categories.
Data events are weighted with fake factors according to the loose-lepton multiplicity of the region:
\begin{equation}
N^{\textrm{fake}} = [ F(N_{TL} + N_{LT}) - F^{2} N_{LL} ]_{\textrm{data}}  - [ F(N_{TL} + N_{LT}) - F^2 N_{LL} ]_{\textrm{MC}}^{\textrm{prompt $\ell$ only}}
\end{equation}
with $N_{TL}$, $N_{LT}$, $N_{LL}$ denoting the number of events in the corresponding category.
The prompt lepton contribution in the events with \texttt{Loose}  leptons is subtracted using the MC simulation to account for the prompt-lepton contamination in the given category.
 
The composition of \rCRSS\ is dibosons (32\%), charge mis-identified background (45\%) and fakes (20\%) in the electron channel, and dibosons (65\%), fakes (27\%) and charge mis-identified background (8\%) in the muon channel.



\section{Boosted channel: event selection and background estimate}

\subsection{Event reconstruction and selection}
\label{subsec:event_selection:boosted}

The boosted analysis is designed for signals in the $\Delta m > 4~\TeV$ region by using the large-$R$ jet to reconstruct the hadrons from the \NR\ decay.
The signal events have a very energetic charged lepton and \NR\ from the \WR\ decay produced back-to-back in the $x$--$y$ plane.
In the highest $\Delta m$ region, the large-$R$ jet overlaps with the lepton from the \NR\ decay.
In the muon channel, such a muon inside the jet can be identified, and a single \texttt{SR} (\bSRTwomu) with two baseline muons is used to cover the full $\Delta m$ range.
On the other hand, in the electron channel, calorimeter clusters from the electron from the \NR\ decay overlap with other hadron activities, making its identification challenging.
Two signal regions, \bSROnee\ and \bSRTwoe, are thus defined requiring exactly one and two baseline electrons in the event, respectively, to cover the higher and lower $\Delta m$ ranges.
The \WR\ mass is reconstructed from the large-$R$ jet and the electron ($m_{eJ}$) in \bSROnee, and from the large-$R$ jet and the two charged leptons ($m_{eeJ}$ and $m_{\mu\mu J}$, or collectively $m_{\ell\ell J}$) in \bSRTwoe\ and \bSRTwomu.
Due to the relatively small background after the event selections, the sensitivity gain by separating \texttt{SR} into same- and opposite-sign events is marginal,
hence the lepton charge is not considered in the boosted analysis.
 
For all three \texttt{SR}s, the requirement that only one large-$R$ jet is reconstructed in the event is imposed, which is expected to cover about 90\% of signal events. This selection cut is useful in suppressing the multijet background,
and improving the fake-lepton, multijet and $\gamma$+jets modelling by the LO simulation with the exception of an overall normalisation factor.
Assuming well-isolated and highly-boosted leptons from the prompt \WR\ decay,
tighter identification and isolation criteria (as detailed in Section~\ref{sec:object_definition}) are
required for the lepton with the highest-\pt\ in the event.
The azimuthal difference between the large-$R$ jet and the highest-\pt\ lepton is required to be greater than 2.
In addition, to suppress the top-quark backgrounds, the number of $b$-tagged small-$R$ jets is required to be zero.
 
\subsubsection{One-electron category (\bSROnee)}
Region \bSROnee\ requires exactly one electron from the \WR\ decay, with the large-$R$ jet encapsulating all decay products of the heavy neutrino $\NR \to eqq^{\prime}$.
A $\met < 200~\GeV$ selection is applied to suppress the main background process, $W (\to e\nu)$ + jets.
The high-\pt\ electron and low \met\ requirements bias the $W$+jets events towards  higher helicity angles, $|\cos\theta| \sim 1.0$, where $\theta$ is the polar angle of the electron from the $W$ boson decay in the rest frame of the $W$\footnote{The $W$ boson rest frame is calculated from the \met and the reconstructed electron by solving the kinematic equation for the neutrino momentum in the $z$-axis direction ($p_{z,\nu}$) and by assuming a $W$ boson mass of 80.4~\GeV.}.
The $|\cos\theta| > 0.7$ region is used for the signal search.
The $|\cos\theta| < 0.7 $ region is used for the normalisation of the $W$+jets contribution.
To further reduce the $\gamma$+jets and dijet contributions, a selection on the $\eta$ difference between the large-$R$ jet and the electron candidate, $\Delta \eta = \left| \eta(J) - \eta(e) \right| < 2$ is applied. The \bSROnee\ definition is summarised in Table~\ref{tab:event_selection:boosted:definition}.
 
\subsubsection{Two-electrons category (\bSRTwoe)}
Region \bSRTwoe\ covers the intermediate $\Delta m$ region between \bSROnee\ and the resolved analysis.
The events are selected if they contain exactly two electrons and one large-$R$ jet.
The  $\met < 200~\GeV$ selection is applied to further reduce the \ttbar\ background.
The dominant background source is $Z (\to ee)$ + jets.
To remove contributions from the pole mass of the $Z$ boson, a $m_{ee} > 200~\GeV$ selection is applied.
The \bSRTwoe\ definition is summarised in Table~\ref{tab:event_selection:boosted:definition}.
\bSROnee\ and \bSRTwoe\ are orthogonal to each other and are statistically combined to obtain the final results.
 
\begin{table}[htb]
\centering
\caption{Definition of the signal regions in the boosted analysis. The baseline definitions of electrons and muons are used. Two regions in the electron channel cover higher- and lower-$\Delta m$ regions, where $\Delta m$ denotes the mass difference between \WR\ and \NR.}
\label{tab:event_selection:boosted:definition}
\begin{tabular}{lccc}
\hline
Region                 & \bSROnee & \bSRTwoe & \bSRTwomu \\
& (higher $\Delta m$) & (lower $\Delta m$) & \\
\hline
Number of large-$R$ jets & \multicolumn{3}{c}{1} \\
\hline
Number of electrons      & 1    &   2       &   0 \\
\hline
Number of muons          & 0    &   0       &   2 \\
\hline
Leading lepton $\pt$ [\GeV]    & \multicolumn{3}{c}{$>200$} \\
\hline
\met[\GeV]                   & \multicolumn{2}{c}{$<200$} & --- \\
\hline
$\left| \cos\theta \right|$             & $> 0.7$ &   ---       &   --- \\
\hline
$\Delta \phi_{J,\ell_1}$   & \multicolumn{3}{c}{$>2.0$} \\
\hline
$\Delta \eta_{J,\ell_1}$   & $<2.0$ &   ---       &   --- \\
\hline
Dilepton $\pt$ (\GeV)        & ---    &   ---       & $>200$ \\
\hline
Dilepton mass $m_{\ell\ell} $[\GeV]       & ---    &   \multicolumn{2}{c}{$>200$} \\
\hline
Number of $b$-tagged small-$R$ jets  & \multicolumn{3}{c}{0} \\
\hline
\end{tabular}
\end{table}
 
\subsubsection{Two-muons category (\bSRTwomu)}
Events are selected in \bSRTwomu\ if they contain exactly two muons and one large-$R$ jet.
A $m_{\mu\mu} > 200~\GeV$ selection is applied.
The upper cut on \met\ is not useful in this channel due to the deteriorated momentum resolution for very high-\pt\ muons.
Instead, the dimuon system's \pt\ is required to be greater than 200~\GeV\ to suppress the \ttbar\ background.
The \bSRTwomu\ definition is also summarised in Table~\ref{tab:event_selection:boosted:definition}.

\subsubsection{Event categorisation for the background estimation}
The normalisations of the main background sources, $W(\to e\nu)$+jets and $Z(\to\ell\ell)$+jets,
as well as for the processes with less reliable yield predictions, $\gamma$+jets and multijet,
are estimated from data in control regions and adjusted by using the \texttt{SR}-to-\texttt{CR} ratio in the simulation.
 
In each category, the reconstructed \WR\ mass distribution is segmented into four bins;
1--2~\TeV, 2--3~\TeV, 3--4~\TeV, and $>4~\TeV$, having considered the search sensitivity for signals with various \WR\ masses and the stability of the background estimation.
High-mass \WR\ signals are expected to be observed in the third and fourth bins,
which are designated bins 1 and 2 of the \texttt{SR} for each channel (for example, bins 1 and 2 of \bSRTwomu\ for the two-muon case).
The first bin of the \WR\ mass histogram is used for the SM background determination in a region where the presence of lower-mass \WR\ signals is negligibly small, given exclusion limits obtained in previous ATLAS analyses~\cite{EXOT-2018-15}.
For the two-electron and two-muon categories, the first bin is used to estimate the $Z$+jets background normalisation in designated \bCRZTwoe\ and \bCRZTwomu, respectively.
In the one-electron case, the 1--2~\TeV\ \mWR\ bin is used to distinguish between $\gamma$+jets and $W$+jets events.
If there is an additional ID track with opposite charge within $\Delta R<0.3$ and $\left| \Delta z_{0} \sin\theta \right| < 0.5$ from the electron candidate track,
the electron is possibly coming from a photon conversion and the events are categorised into \bCRyOnee.
The remaining events are categorised into \bCRLowOnee.
The fractions of $W$+jets, $Z$+jets, multijet and $\gamma$+jets are about 33\%, 10\%, 24\% and 30\%, respectively, in \bCRyOnee,
and approximately 73\%, 17\%, 3\% and 1\% in \bCRLowOnee.
 
The normalisation for $W(\to e\nu)$+jets and multijet events is estimated in the dedicated control regions, \bCRWOnee\ and \bCRFakeOnee,
requiring $|\cos\theta| < 0.7$ and the inverted electron isolation cut, respectively, with the remaining selection identical to that of \bCRLowOnee.
The purity of the $W$+jets and multijet events in \bCRWOnee\ and \bCRFakeOnee\ is about 83\% and 91\%.
One should note that the $\gamma$+jets contribution in \bCRFakeOnee\ is not large
since mis-identified electrons from photon conversions can satisfy the tight isolation requirement.
 
The estimated \texttt{CR} normalisation factors are extrapolated to higher-mass bins.
The validity of the method is first checked in the second bin ($\mWR = [2, 3]~\TeV$) in each channel: \bVRTwoe, \bVRTwomu\ and \bVROnee, for \bSRTwoe, \bSRTwomu\ and \bSROnee, before the third and fourth bins are tested.
In addition, the regions with $\left|\cos\theta\right| <0.7$ and the inverted electron isolation in the one-electron channel
are used to check the extrapolation of $W$+jets and multijet normalisations from the lower to the higher \WR\ mass bins.
The regions are segmented into 2--3~TeV and $>3~\TeV$ and labelled as \bVRWOnee\ and \bVRFakeOnee, respectively.
 
To check the extrapolation of the $Z$+jets normalisation from the lower to the higher  \WR\ mass bins, additional validation regions are defined (\bVRZTwoe\ and \bVRZTwomu), but which are not used in the final fit.
They require $120 < m_{\ell\ell} < 200~\GeV$ instead of $>200~\GeV$.
 
The definitions of \texttt{SR}s, \texttt{CR}s and \texttt{VR}s for the boosted channel are visualised in Figure~\ref{fig:background_estimation:boosted:region_definition}. Validation results are presented in Sec.~\ref{results:boosted}.
 
\begin{figure}[htb]
\centering
\subfloat[]{
\includegraphics[width=0.47\textwidth]{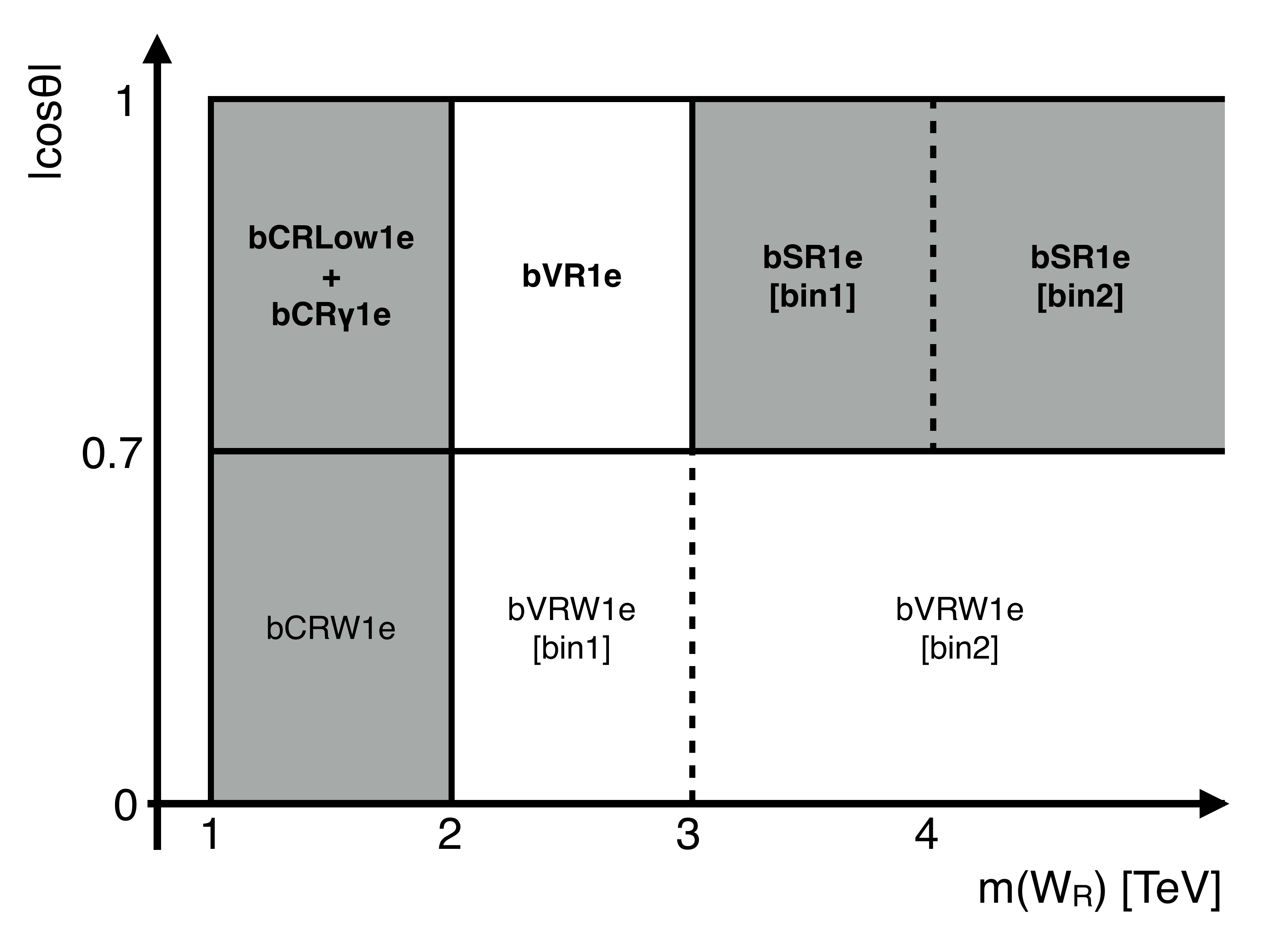}
}
\subfloat[]{
\includegraphics[width=0.47\textwidth]{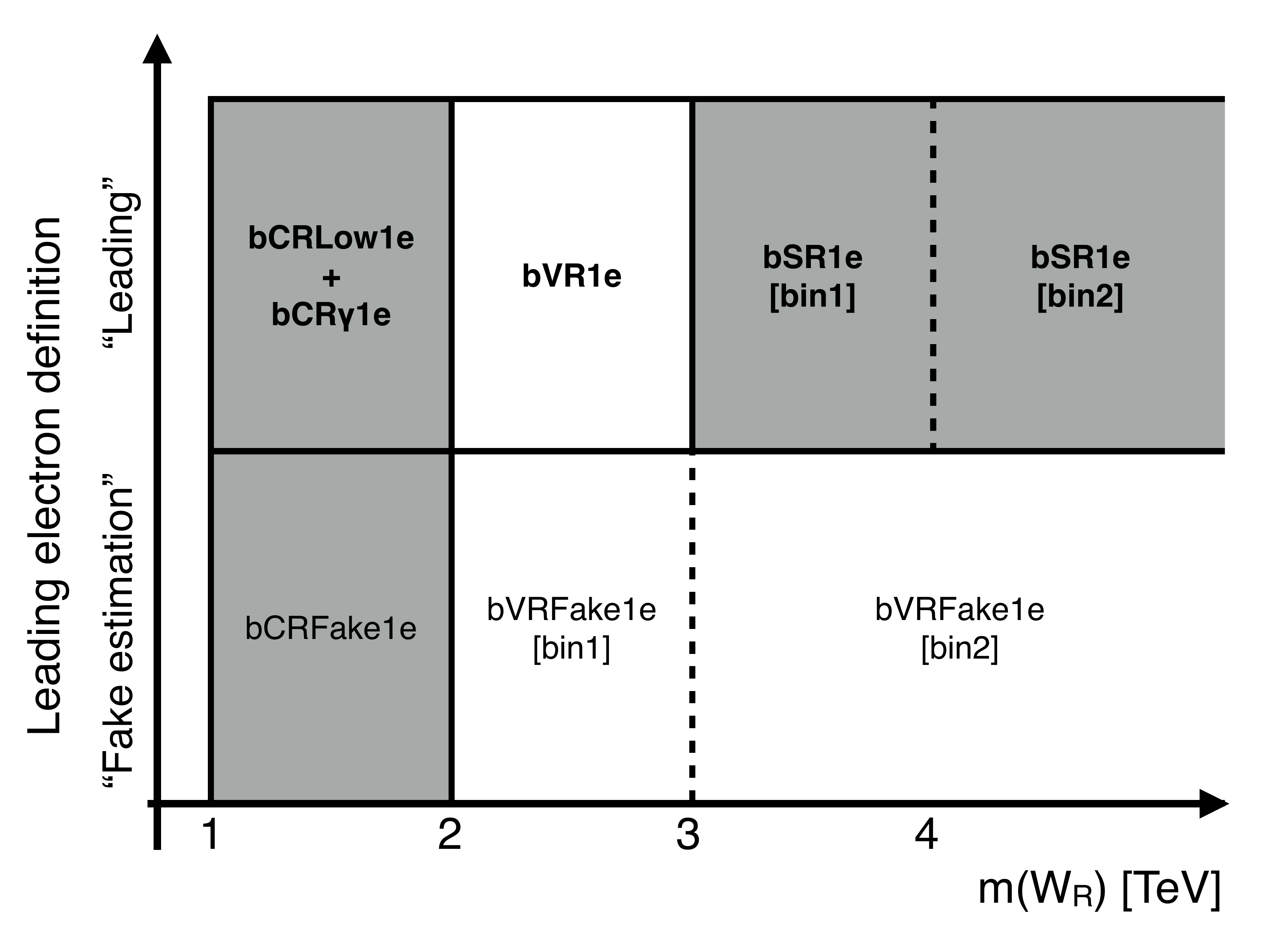}
}
\\
\subfloat[]{
\includegraphics[width=0.47\textwidth]{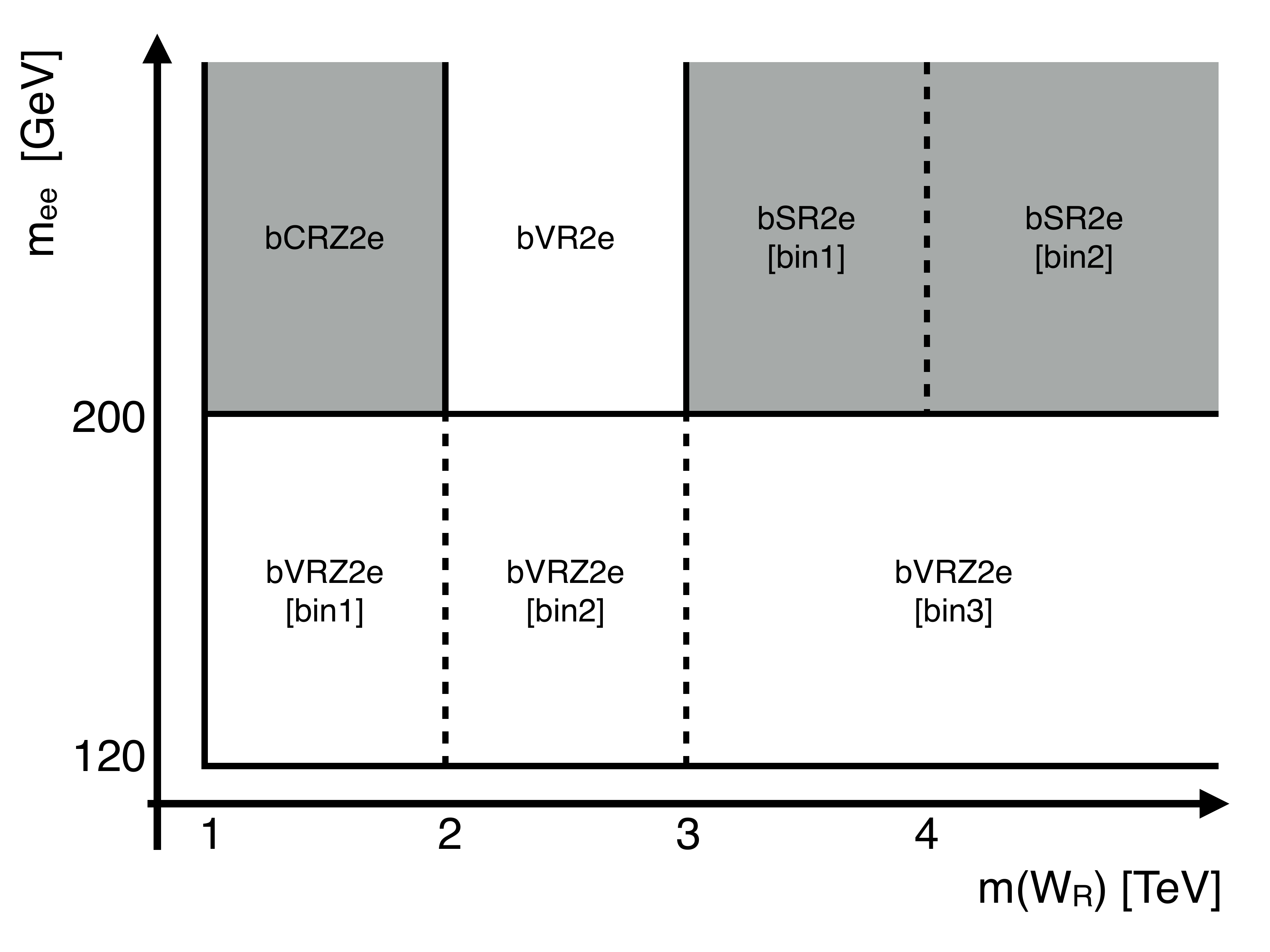}
}
\subfloat[]{
\includegraphics[width=0.47\textwidth]{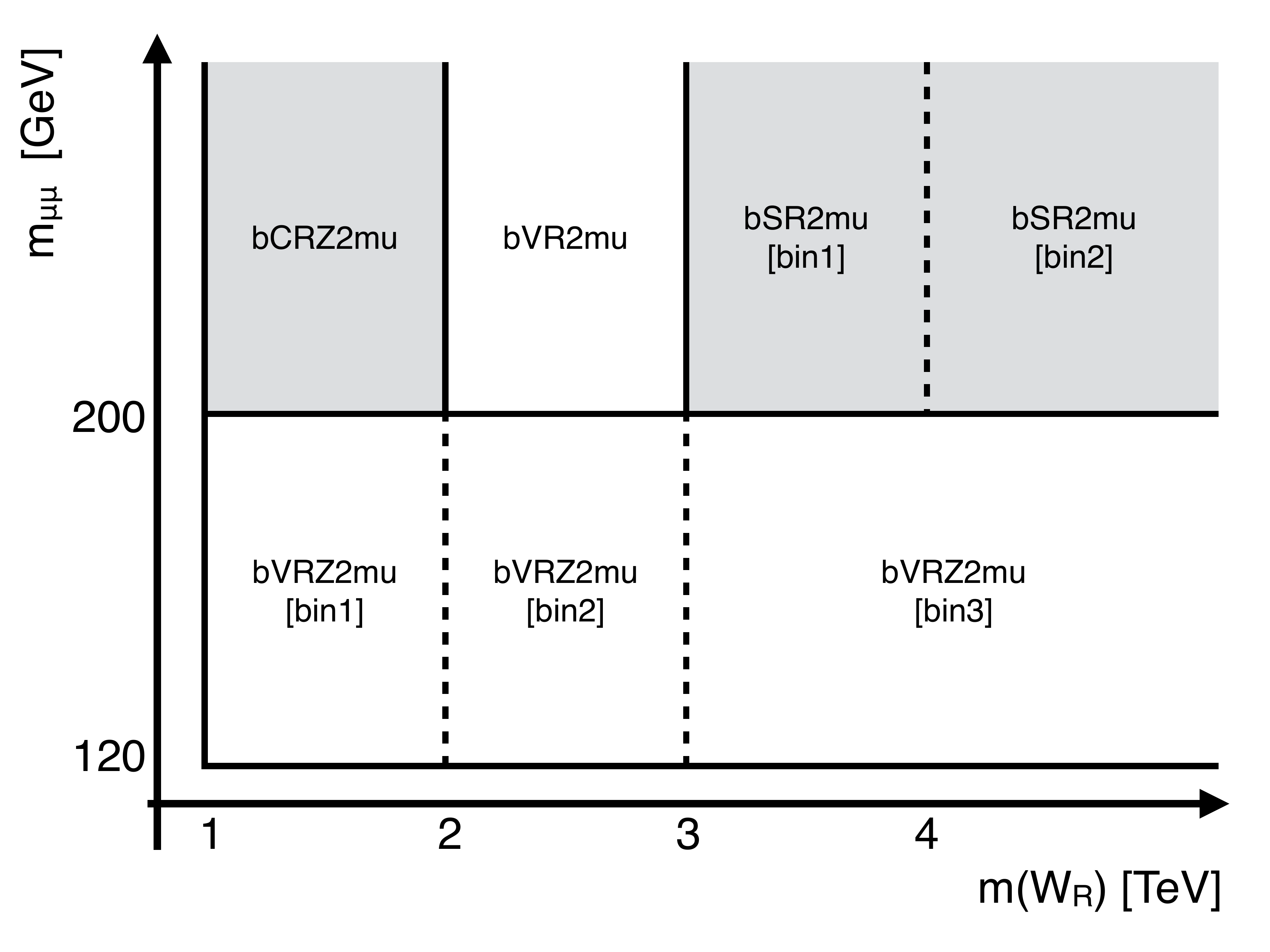}
}
\\
\caption{
Boosted channel: Schematic view of \texttt{SR}/\texttt{CR}/\texttt{VR}s definitions for \bSROnee\ (a)--(b), \bSRTwoe\ (c) and \bSRTwomu\ (d).
(a) and (b) show the regions for the estimation of $W$+jets and multijet backgrounds, respectively. The labels
\bSROnee, \bVROnee, \bCRLowOnee\ and \bCRyOnee\ shown in bold in the two figures correspond to the same regions.
The dark-grey coloured regions in (a), (b) and (c) are contributing simultaneously to the final fit in the electron channel,
while the light-grey coloured regions in (d) are used for the muon channel, separately.
The remaining regions are used to verify the background estimations (see text for details).
}
\label{fig:background_estimation:boosted:region_definition}
\end{figure}


\subsection{Background estimation}
\label{subsec:background_estimation:boosted}

For the $W$+jets and $Z$+jets estimation
in the boosted analysis, since the number of jets is forced to be one,
the correction for jet multiplicity mis-modelling used in the resolved analysis is not required.
The simulated \WR\ mass distributions for $W$+jets in \bVRWOnee\ (with $|\cos\theta|<0.7)$ and those for $Z$+jets \bVRZTwoe\ and \bVRZTwomu\ agree well with the observed data.
The \WR\ mass distributions in these regions are only used to confirm
the normalisation methodology for each process and these are within the considered uncertainties.
 
The fake lepton contribution is negligibly small in \bSRTwoe\ and \bSRTwomu\ after the requirement of a high-\pt\ selection on the leading lepton.
It is, however, not negligible at the higher \WR\ mass tail in \bSROnee.
MC simulation is used to estimate the non-prompt electron contribution with a data-driven correction factor obtained in the \texttt{CR}.
The simulated \WR\ mass in \bVRFakeOnee\ agrees well with the multijet sample within the uncertainties considered.
In the final fits the normalisation is mainly constrained by \bCRFakeOnee\ and its validity in the \texttt{SR} relies on the good
simulation of the isolation efficiency for fake electrons.
To check the validity of this approach, two additional regions
beyond those shown in Figure~\ref{fig:background_estimation:boosted:region_definition} are defined, corresponding to \bSROnee\ and \bCRFakeOnee:
the electron candidate is forced to satisfy the \texttt{Medium} identification but fail the \texttt{Tight}  identification criterion in order to be orthogonal to the \texttt{SR}, \texttt{CR}s and \texttt{VR}.
The remaining selection is the same as in \bSROnee\ (\bCRFakeOnee), with the requirement to satisfy (fail) the \texttt{HighPtCaloOnly} isolation criterion.
In both regions, the purity of the multijet process is greater than 90\%.
By comparing data to MC in both regions, it is found that the simulated isolation efficiency for fake electrons agrees with the data within the considered uncertainties.
A small data/MC disagreement on \mWR\ distribution is considered as an additional uncertainty on the multijet estimation in \bSROnee.
 
The $\gamma$+jets contribution with isolated electrons from photon conversions is also estimated by MC simulation with a normalisation factor determined in a data-driven way.
Additional validation regions are defined requiring exactly one reconstructed photon instead of an electron, with the remaining selection the same as in \bSROnee.
The purity of $\gamma$+jets sample in this region is about 70\%, and the fraction of multijet events is 30\%.
The reconstructed \WR\ mass with the photon (in the place of the electron) and the large-$R$ jet agrees between data and MC within the considered uncertainties.
The \pt- and $\eta$-dependent photon conversion rate may change the reconstructed \WR\ mass distribution for the $\gamma$+jets process, but this rate mainly depends on the detector geometry and it is well modelled in the simulation.
The normalisation of the $\gamma$+jets contribution is estimated mainly from \bCRyOnee, by requiring an additional ID track close to the electron.
The modelling of the fraction of events with an ID track close to the electron in simulated $W$+jets and multijets can affect the $\gamma$+jets estimation, so a comparison is carried out
with data in \bCRWOnee\ and \bCRFakeOnee.
For the $\gamma$+jets contribution, the same check is performed in the $\Delta \eta > 2.0$ region.
The purity of $\gamma$+jets in this region is about 50\%, and a conservative uncertainty on the other background subtraction is included in the study.



\section{Systematic uncertainties}
\label{sec:systematics}

 
Several experimental and theoretical systematic uncertainties are considered, affecting both the background and signal predictions as well as the total event yield.
In addition, the statistical uncertainty of the MC simulated events is also considered.
These uncertainties also affect the shape of the variables used in the fit, with the exception of the luminosity and cross-section uncertainties.
 
The uncertainty on the integrated luminosity is 1.7\%~\cite{ATLAS-CONF-2019-021},
It is obtained using the LUCID-2 detector~\cite{avoni2018new} for the primary luminosity measurements.
The uncertainty due to the pile-up reweighting procedure is estimated by varying the amount of pile-up in the simulation
to cover the uncertainty in the ratio of the predicted and measured inelastic cross section~\cite{STDM-2015-05}.
 
A set of experimental systematic uncertainties arise from the energy and momentum calibrations of leptons and jets, the lepton reconstruction, isolation and trigger efficiencies, and the jet vertex tagger and flavour-tagging efficiencies.
The largest uncertainty in the total SM yield arises from the energy calibration and smearing of
jets, derived in Ref.~\cite{JETM-2018-05}, and is between 1 and 6\% in the SS and about 4\% in the OS in the resolved channel, and between 5 and 12\% in the boosted channel, depending on the signal region.
Uncertainties associated with lepton reconstruction, identification,
isolation and trigger efficiencies, as well as energy or momentum calibration~\cite{EGAM-2018-01, TRIG-2018-05, MUON-2018-03} and $b$-jet tagging~\cite{FTAG-2018-01, FTAG-2020-08}, vary
between 5\% (OS) and 13\% (SS) of the total SM yield in the resolved analysis, and between 2\% and 5\% in the boosted.
These uncertainties are propagated to the \met calculation.
In addition, uncertainties on the scale and resolution of the `soft term' of the \met\ are taken into account~\cite{PERF-2016-07}.
In \bSRTwomu, the muon \pt\ is further smeared by taking into account the impact of MS outliers on the muon resolution not described by the simulation. This additional effect is considered as the high-\pt\ muon uncertainty.
 
There are three additional sources of systematic uncertainty associated with the background estimation techniques.
In \rSRSS, the uncertainty related to the charge mis-identification probability of electrons arises from the statistical uncertainty of  the data and simulated samples of $Z$+jets events used in this measurement. The uncertainty is around 6.8\% with a mild dependence on the electron \et\ and $\eta$~\cite{Aaboud_2019}.
The uncertainty on the fake estimation arises from the limited knowledge in the composition of fakes, as well as from the statistical uncertainty
and prompt lepton subtraction used to derive $F$ in the fake-enriched regions.
The uncertainty due to the composition of fakes is estimated by varying the nominal criteria defining the selection sample used in the
fake-factor measurement~\cite{EXOT-2018-33}. The effect on the SM yields is 3.8\% (electrons) and 0.8\% (muons).
In \rSROS, an additional uncertainty is associated with the $m_{jj}$ reweighting of the
$Z$+jets process. This is evaluated by comparing the shape difference between the reweighted simulated $m_{jj}$ distribution and the one measured in data, using \rVROS: the
uncertainty in the reweighting factor is found to vary between 0.4\% and 0.8\%, as a function of the dijet invariant mass.
This observed difference covers all possible mismodellings of the $Z$+jets shape, hence additional theoretical uncertainties for the $Z$+jets process are not considered in the resolved channel.
 
In the boosted channel,
the theory uncertainties estimated for the $W$+jets, $Z$+jets, diboson, multijet and $\gamma$+jets background processes are considered,
which include the choice of QCD renormalisation ($\mu_r$) and factorisation ($\mu_f$) scales, choice of the PDF set and $\alpha_S$, as well as
CKKW matching scale and the resummation scale.
The QCD scale uncertainty is estimated by varying $\mu_r$ and $\mu_f$ to half and twice their nominal values. The PDF uncertainty is estimated using the envelope of the NNPDF3.0 PDF
set, as recommended in Ref.~\cite{pdf4lhc}. In addition, the MMHT2014~\cite{Harland_Lang_2015} and CT14NNLO~\cite{PhysRevD.93.033006} PDF
sets are used to estimate the uncertainty due to the PDF choice. Moreover, the uncertainty due to
$\alpha_S$ is evaluated by varying its nominal value of 0.118 by $\pm 0.001$.
The CKKW matching scale is the parameter to remove the overlap between jets from the matrix element calculation and the parton shower algorithm.
The nominal value is 20~GeV, and the uncertainty is estimated by varying it to 30~GeV and 15~GeV.
The uncertainty on the resummation scale of soft gluon emission is estimated by varying the parameter to half and twice the nominal value.
The largest theory uncertainties generally originate from the CKKW matching and \texttt{Sherpa} resummation scales ($\sim$ 20\%), and QCD scales variations which range between 10\% and 20\%, depending on
the simulated process and the mass of the target signal.
 
The theory uncertainties associated with \ttbar\ processes are as follows.
The uncertainty from hard scatter generation is evaluated by comparing the \texttt{Powheg-Box} and \texttt{MG5\_aMC@NLO} generators,
both interfaced to the \texttt{Pythia}~8.186~\cite{Sj_strand_2008} parton shower model. The
uncertainty due to the hadronisation and fragmentation model is determined by comparing the
nominal \texttt{Powheg-Box} + \texttt{Pythia}~8.186 generated sample with the one generated by \texttt{Powheg-Box}
interfaced to \texttt{Herwig}~\cite{B_hr_2008} (version 7.13). The uncertainty related to the amount of initial- and
final-state radiation (ISR and FSR) is assessed by varying parton shower settings.
The largest theory uncertainty generally originates from the amount of initial- and final-state radiation and is between 0.4\% -- 0.5\% (ISR) and 0.6\% (FSR)
of the \ttbar\ process yield.
 
In the boosted analysis, additional uncertainties are considered by comparing different MC samples
on the possible variations of the reconstructed \mWR, $\cos\theta$ and the isolation efficiency for the fake electrons.
For $W$/$Z$+jets, the nominal \texttt{Sherpa}~2.2.11 samples are compared with the samples generated with \texttt{MadGraph5\_aMC@NLO}+\texttt{Pythia}~8.186.
For $\gamma$+jets, the nominal \texttt{Sherpa}~2.1 sample is compared with the alternative sample generated with \texttt{Pythia}~8.235~\cite{Sjostrand:2014zea}.
The nominal \texttt{Pythia}~8.230 multijet MC sample is compared with the sample generated with \texttt{Sherpa}~2.2.5.
For the multijet sample, an additional uncertainty on the \mWR\ shape is considered by taking the residual non-closure in studies employing different electron identification discussed in Section~\ref{subsec:background_estimation:boosted}.
The possible mis-modelling of the \bCRyOnee\ to \bCRLowOnee\ ratio is evaluated by considering the observed discrepancy in the \texttt{CR}s.
 
The theory uncertainty of the signal efficiency times acceptance amounts to 20\%. It is
evaluated by varying renormalisation and factorisation scales as described above and by using
alternative PDF sets, CTEQ6~\cite{Pumplin_2002} and MSTW~\cite{Martin_2009}. The $\alpha_S$ emission scale factor is also varied to
half and twice the nominal value. The uncertainty is dominated by the variation in factorisation
scale. The variations are performed using SysCalc~\cite{syscal}.
 
The systematic uncertainties in each \texttt{SR} are summarised in Figure~\ref{fig:systs}. Individual uncertainties can be correlated with others, and do not necessarily add in quadrature towards the total background uncertainty.
 
\begin{figure}[phtb]
\centering
\subfloat[]{
\includegraphics[width=0.59\textwidth, keepaspectratio]{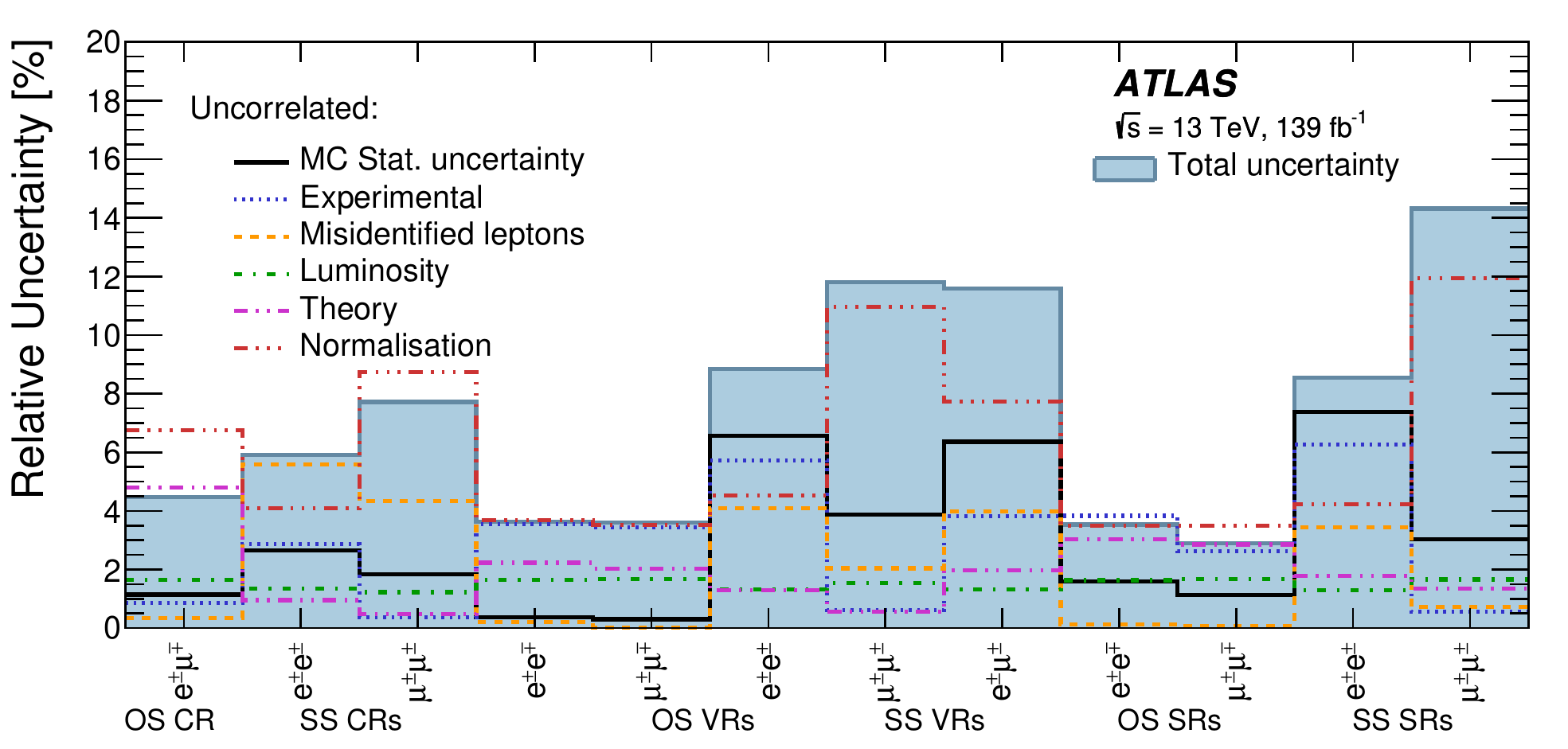}
} \\
\subfloat[]{
\includegraphics[width=0.6\textwidth, keepaspectratio]{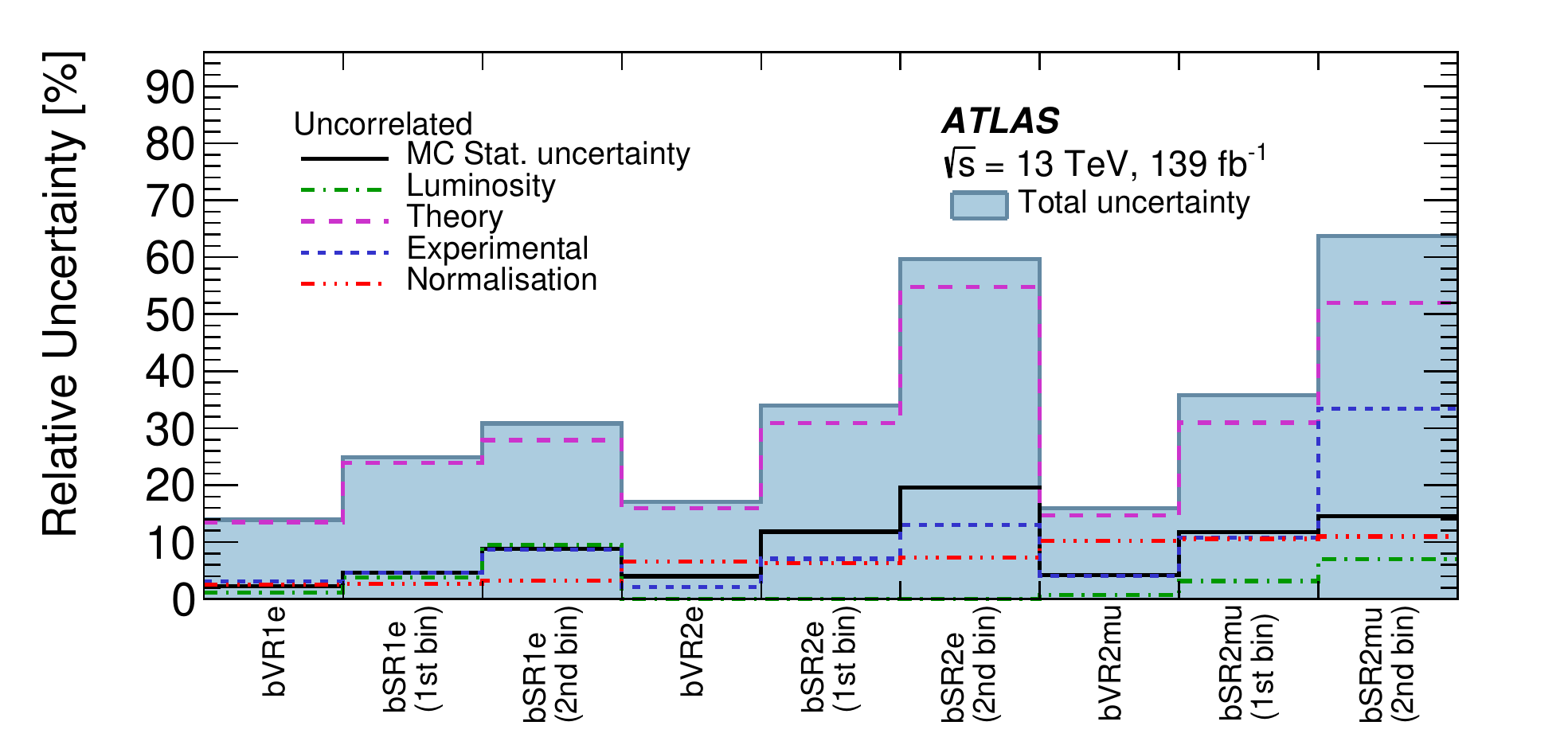}
}
\caption{Relative uncertainties in the total background yield post-fit estimates for the (a) resolved and (b) boosted channels.
``Theory'' indicates the theoretical uncertainty associated with the simulated physics processes (e.g. cross sections).
``Normalisation'' is the uncertainty associated with the yield variations of the dominant backgrounds in the fit.
``Experimental'' corresponds to the combined uncertainty on physics object efficiencies (such as trigger, identification or isolation) and uncertainties associated with \met\ and pile-up. The total uncertainty takes into account any correlations among nuisance parameters.
}
\label{fig:systs}
\end{figure}


\section{Statistical analysis and results}
\label{sec:results}

The search for a \WR\ and \NR\ signal is carried out using the theoretical model described in Section~\ref{sec:signal_model}. The statistical treatment is done separately for samples containing electrons and muons since there is no theoretical motivation that the \NR\ should have the same mass across different flavours. The resolved and boosted channels are analysed independently. The statistical interpretation is performed by carrying out binned maximum-likelihood fits using the HistFitter~\cite{Baak:2014wma} framework. The likelihood is a product of Poisson probability density functions, describing the observed number of events in each bin of regions involved in the fit, and Gaussian distributions that describe the nuisance parameters associated with each of the systematic uncertainties. Systematic uncertainties that are correlated between different samples and different regions are accounted for via a common nuisance parameter.
 
A ``background-only'' fit is carried out first by using the observed event yield in \texttt{CR}s assuming that no signal contributes in these regions, and by applying the resulting normalisation factor to the number of background events predicted by simulation for the equivalent
process in the validation
regions. The normalisation factors of the background processes are allowed to vary freely.
The \texttt{VR}s are only employed to confirm the validity of the background modelling in the background-only \texttt{CR} fits, and do not contribute to the search results presented here.
 
A non-zero signal contribution for each $\mWR$\ and $\mNR$ mass point is then allowed involving both \texttt{SR} and \texttt{CR}s in the fit, as summarised in Figures~\ref{fig:background_estimation:resolved:region_definition} and \ref{fig:background_estimation:boosted:region_definition}. The normalisation factors of the various (signal or background) processes are allowed to vary freely.
Upper limits at 95\% C.L. on the signal strength of the KS process are calculated using the CL$_{\text s}$ method~\cite{Read_2002} and the profile likehood-ratio as the test statistic. The asymptotic method \cite{Cowan_2011} was used, but similar results were obtained with toy MC tests.

\subsection{Resolved channel}
\label{results:resolved}
In the resolved channel, the $m_{\ell\ell jj}$ distribution is used in the \rSROS\ fit when $\mWR > \mNR$. In the $\mWR < \mNR$ case, the fit results are obtained by employing the $m_{jj}$ distribution. The $h_{\textrm{T}}$ distribution is used in the \rSRSS\ fit.
For the scenario in which the \NR\ neutrino is a Majorana particle, the OS and SS channels are fitted simultaneously, whereas for the Dirac neutrino scenario, only the OS channel is used in the fit.
 
A combined ``background-only'' fit is first carried out in the \rCROSemu, \rCRSSTwoe\ and \rCRSSTwomu\ regions. The free normalisation factors for the \ttbar\ and diboson processes are estimated mainly from \rCROSemu\ and \rCRSSTwomu.
The validity of the evaluated normalisation factors for same-sign $Z$+jets, \ttbar\ and diboson samples is confirmed by applying them in the \rVROSTwoe, \rVROSTwomu, \rVRSSemu, \rVRSSTwoe, and \rVRSSTwomu\ regions.
All systematic uncertainties described in Section~\ref{sec:systematics} are considered as nuisance parameters.
Figure~\ref{fig:VRdataMC} shows an example of the post-fit $m_{\ell\ell jj}$ distribution in \rVROSTwomu\ and $h_{\textrm{T}}$ distribution in \rVRSSTwoe, showing a good agreement between the observed data and the expected background.
 
\begin{figure}[phtb]
\centering
\subfloat[]{
\includegraphics[width=0.43\textwidth, keepaspectratio]{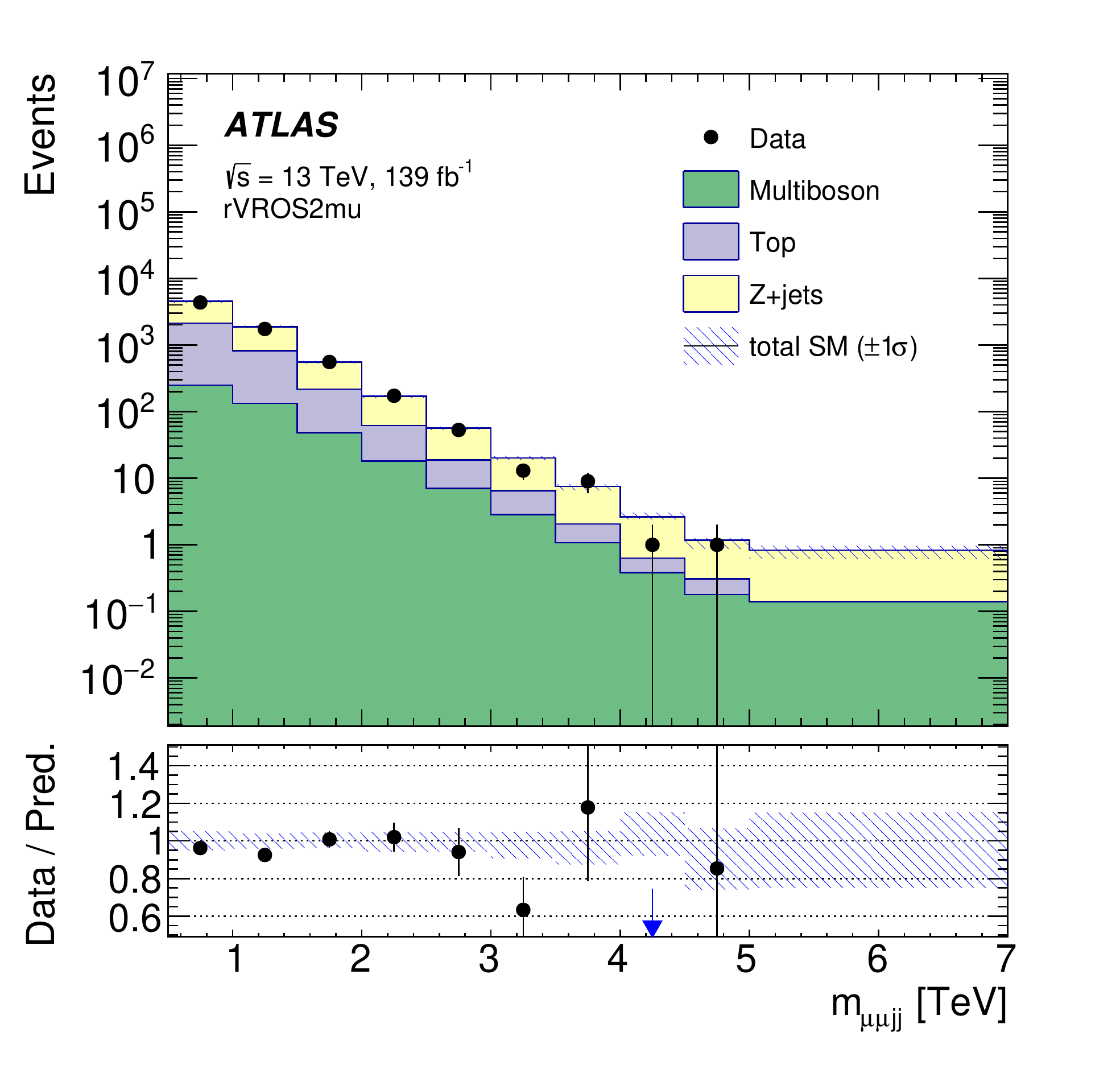}
}
\subfloat[]{
\includegraphics[width=0.43\textwidth, keepaspectratio]{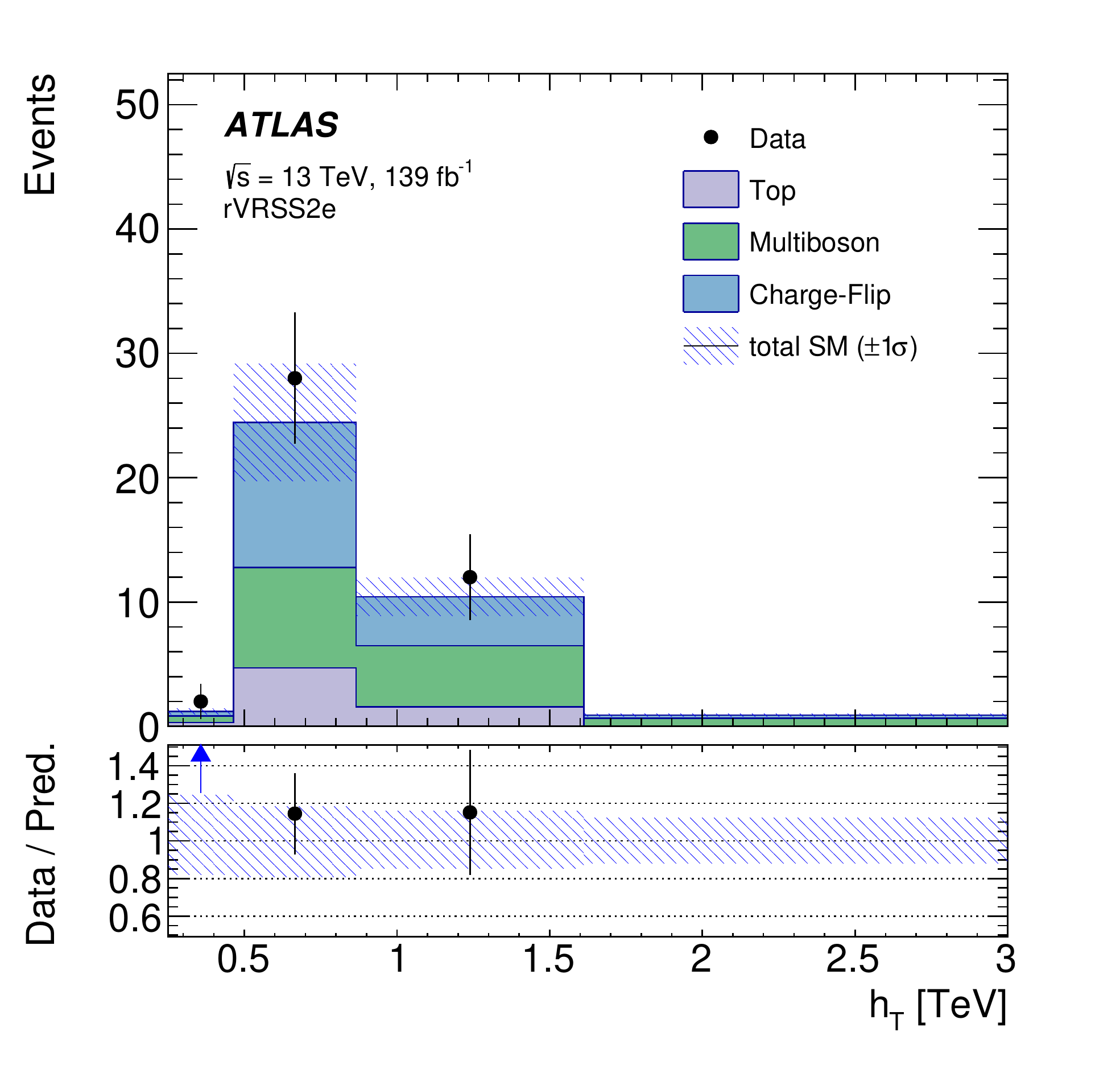}
}
\caption{
\texttt{VR} fits in the resolved channel: (a) $m_{\ell\ell jj}$ in \rVROSTwomu\ and  (b) $h_{\textrm{T}}$ distribution in \rVRSSTwoe.
`Top' refers to all processes containing at least one top quark. `Multiboson' refers to $VV$ and $VVV$ processes, where $V$ = $W$ or $Z$.
The expected background is determined via a fit on the \texttt{CR} data.
The hatched band (`total SM') includes all post-fit systematic uncertainties, having taken into account all correlations among various sources. A blue arrow indicates an out-of-range data point for a given bin.
}
\label{fig:VRdataMC}
\end{figure}
 
The integrated event counts for the observed data and the estimated background for the \texttt{CR}, \texttt{VR} and \texttt{SR} regions  are shown in Figure~\ref{fig:VRpulls}.
The post-fit $m_{\ell\ell jj}$ and $m_{jj}$ distributions  in \rSROS\ and $h_{\textrm{T}}$ distributions in \rSRSS\ are shown in Figure~\ref{fig:SRdataMC_resolved}.
In all cases, no significant deviation above the expected background is observed in the data.
The largest excess is observed in \rSRSS2e\ around $h_T \sim$ 1.0 TeV (Figure \ref{fig:SRdataMC_resolved}e): 39 events are observed in a region where 24.0 $\pm$ 2.4 background events are expected. The observed deviation is in a background-rich region and is not consistent with the shape of the signal.

Exclusion limits at 95\% CL on the signal strength of the KS process as a function of \mWR\ and \mNR\ are calculated.
The results are shown in Figure~\ref{fig:limit}. For electron Majorana neutrinos, the excluded region extends to a \WR\ mass of 5.6~TeV  and to a \NR\ mass of  3.5~TeV. For muon Majorana neutrinos, \WR\ masses of up to 5.7~TeV were excluded, and \NR\ masses of up to 3.6~TeV. For electron Dirac neutrinos, the excluded region extends to a \WR\ mass of 5.9~TeV and to a \NR\ mass of  3.6~TeV. For muon Dirac neutrinos, \WR\ masses of up to 5.8~TeV were excluded, and \NR\ masses of up to 3.8~TeV.
Compared to the previous ATLAS analysis in the resolved channel~\cite{EXOT-2017-18}, the exclusion limits on the \WR\ masses are extended by 0.8--1.2~\TeV\ in the case of $\mWR > \mNR$.
 
\begin{figure}[htb]
\centering
\includegraphics[width=0.8\textwidth, keepaspectratio]{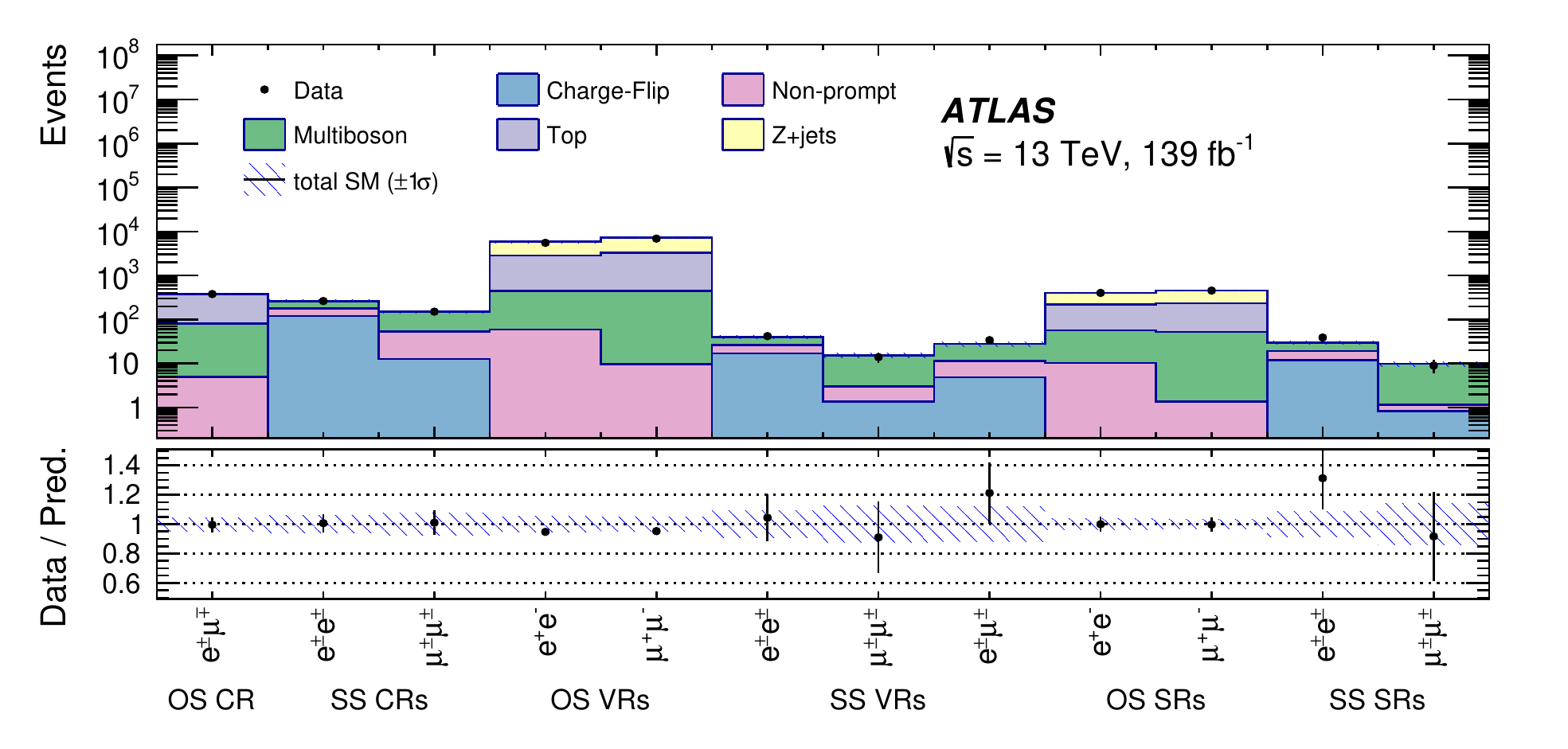}
\caption{\texttt{CR}/\texttt{VR}/\texttt{SR} fits in the resolved channel: Integrated number of events for observed data and expected background in $e\mu$ (OS), and $ee$ and $\mu\mu$ (SS) \texttt{CR}, $ee$ and $\mu\mu$ (OS), and $ee$, $\mu\mu$ and $e\mu$ \texttt{VR} and $ee$ and $\mu\mu$ (both OS and SS) \texttt{SR}s. `Top' refers to all processes containing at least one top quark. `Multiboson' refers to $VV$ and $VVV$ processes, where $V$ = $W$ or $Z$.
The expected background is determined via a fit on the \texttt{CR} data.
The hatched bands (`total SM') include all post-fit systematic uncertainties, having taken into account all correlations among various sources.
}
\label{fig:VRpulls}
\end{figure}
 
\begin{figure}[hptb]
\centering
\subfloat[]{
\includegraphics[width=0.37\textwidth, keepaspectratio]{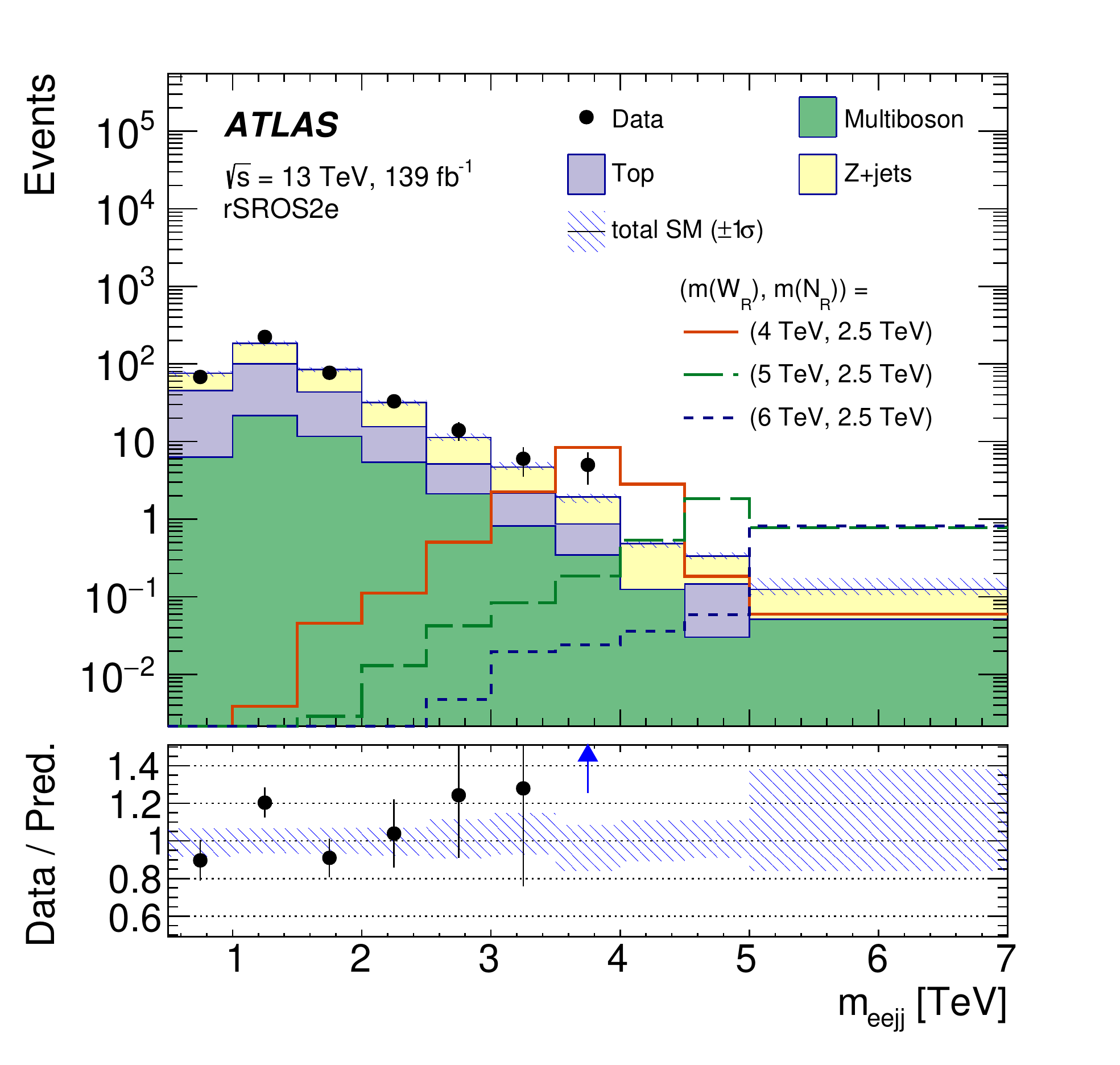}
}
\subfloat[]{
\includegraphics[width=0.37\textwidth, keepaspectratio]{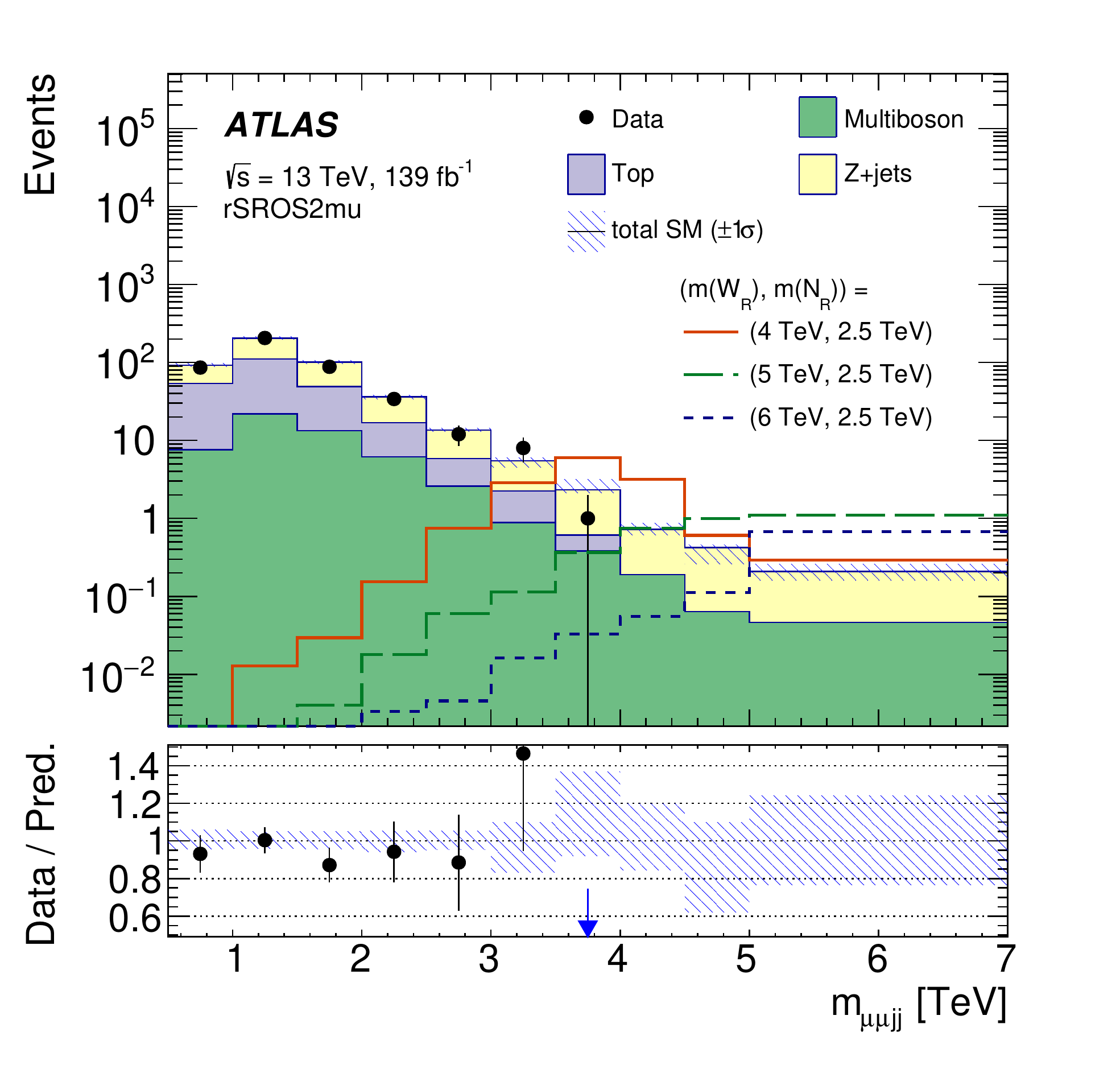}
} \\
\subfloat[]{
\includegraphics[width=0.37\textwidth, keepaspectratio]{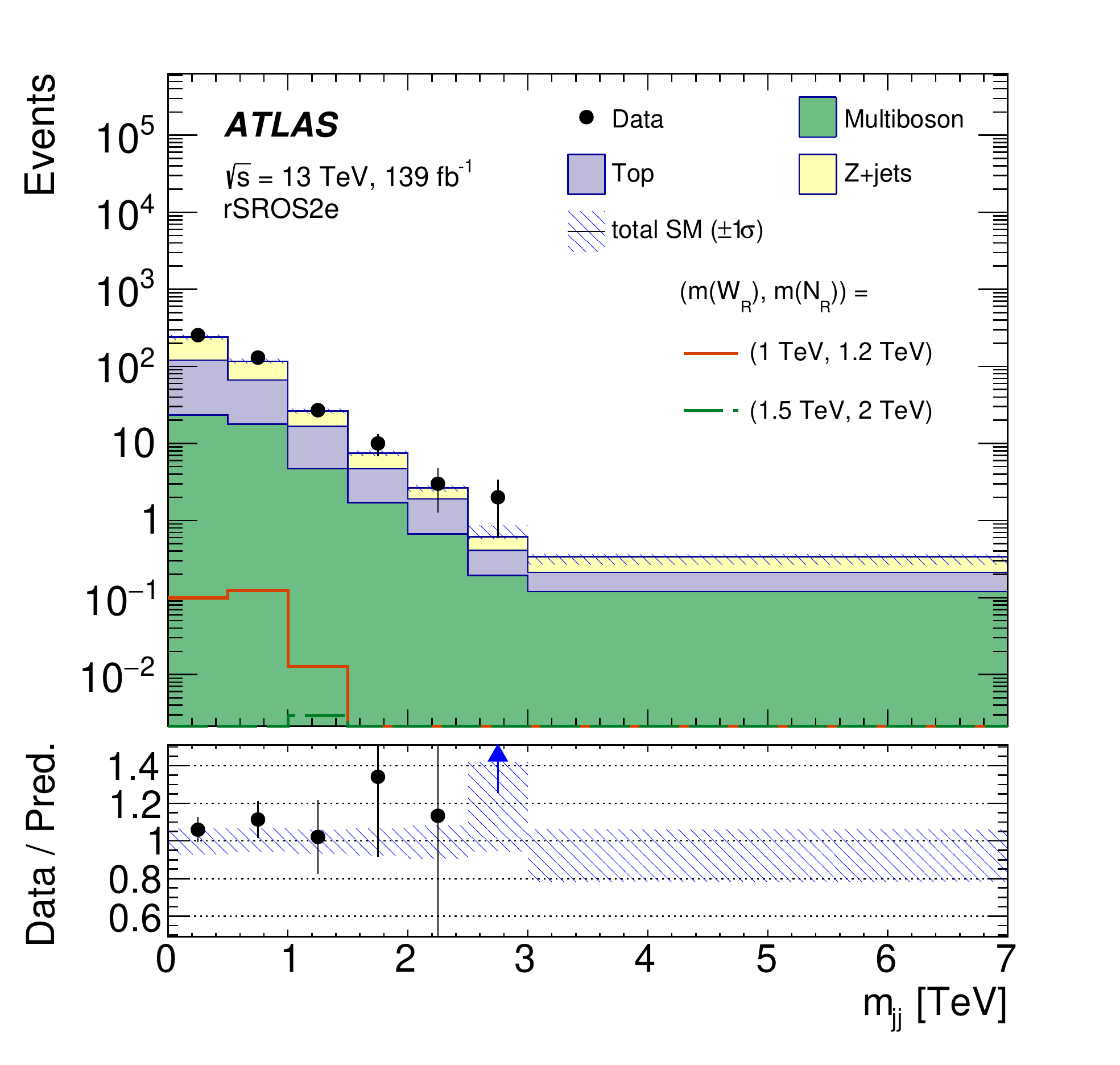}
}
\subfloat[]{
\includegraphics[width=0.37\textwidth, keepaspectratio]{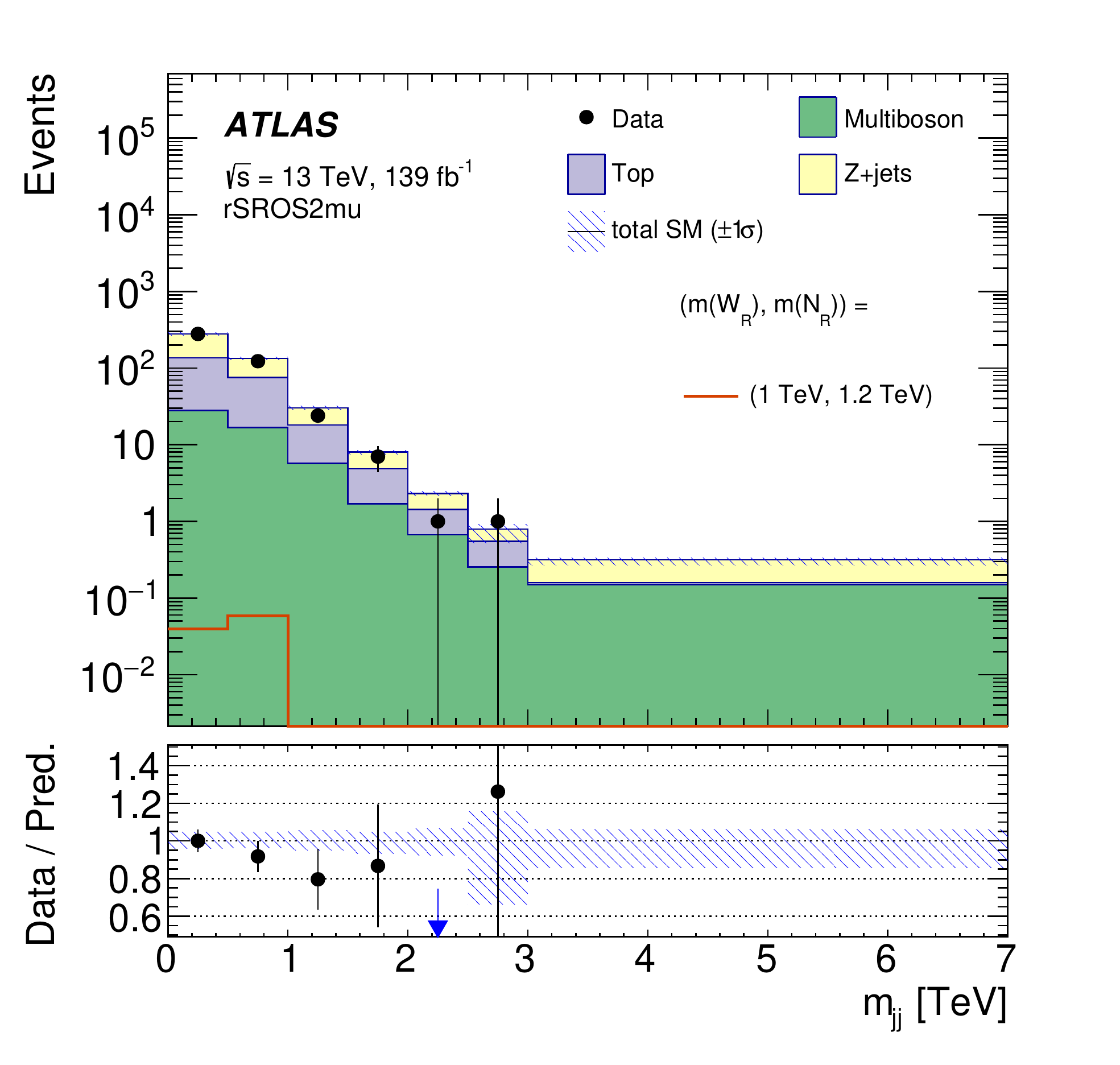}
} \\
\subfloat[]{
\includegraphics[width=0.37\textwidth, keepaspectratio]{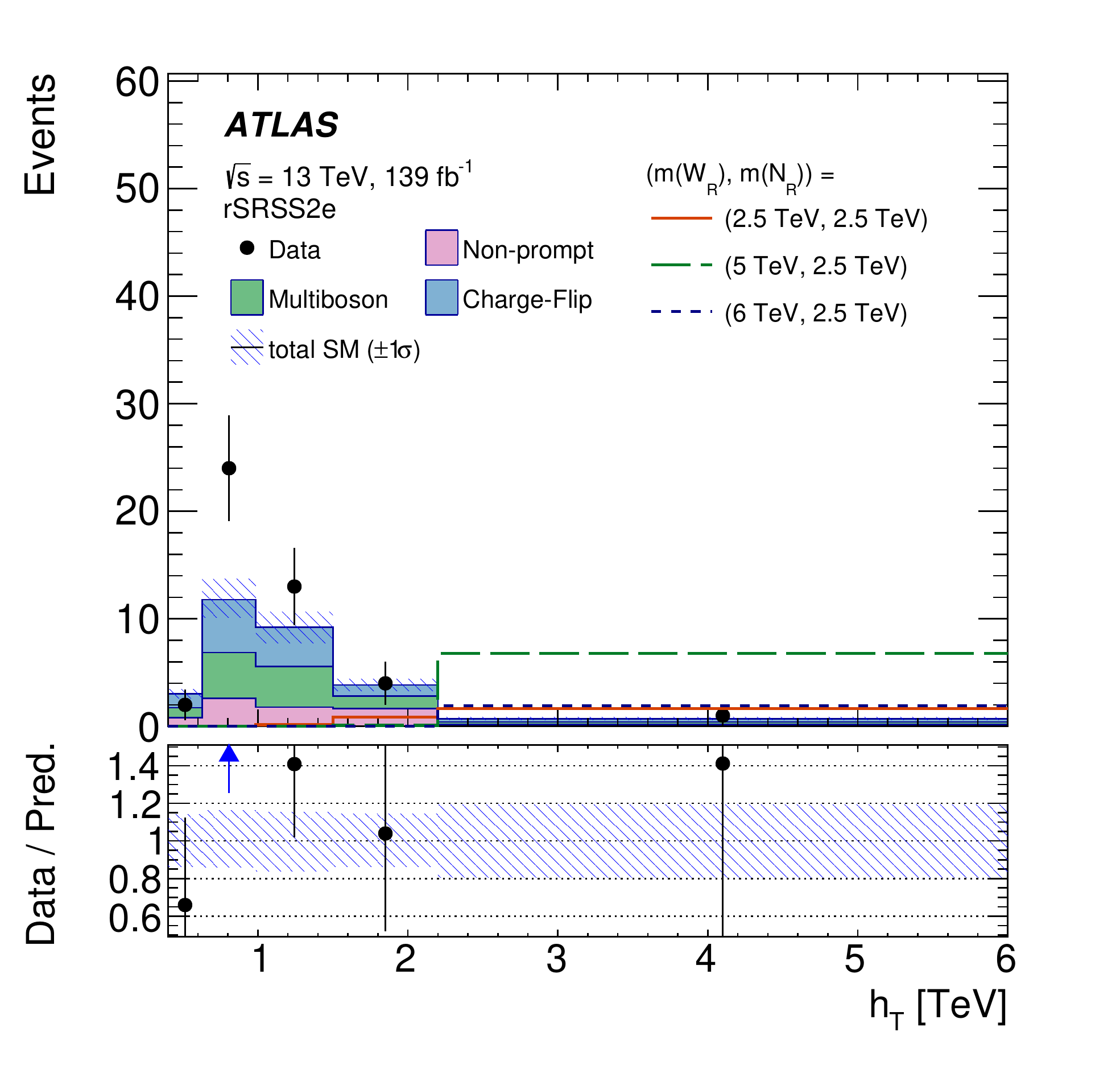}
}
\subfloat[]{
\includegraphics[width=0.37\textwidth, keepaspectratio]{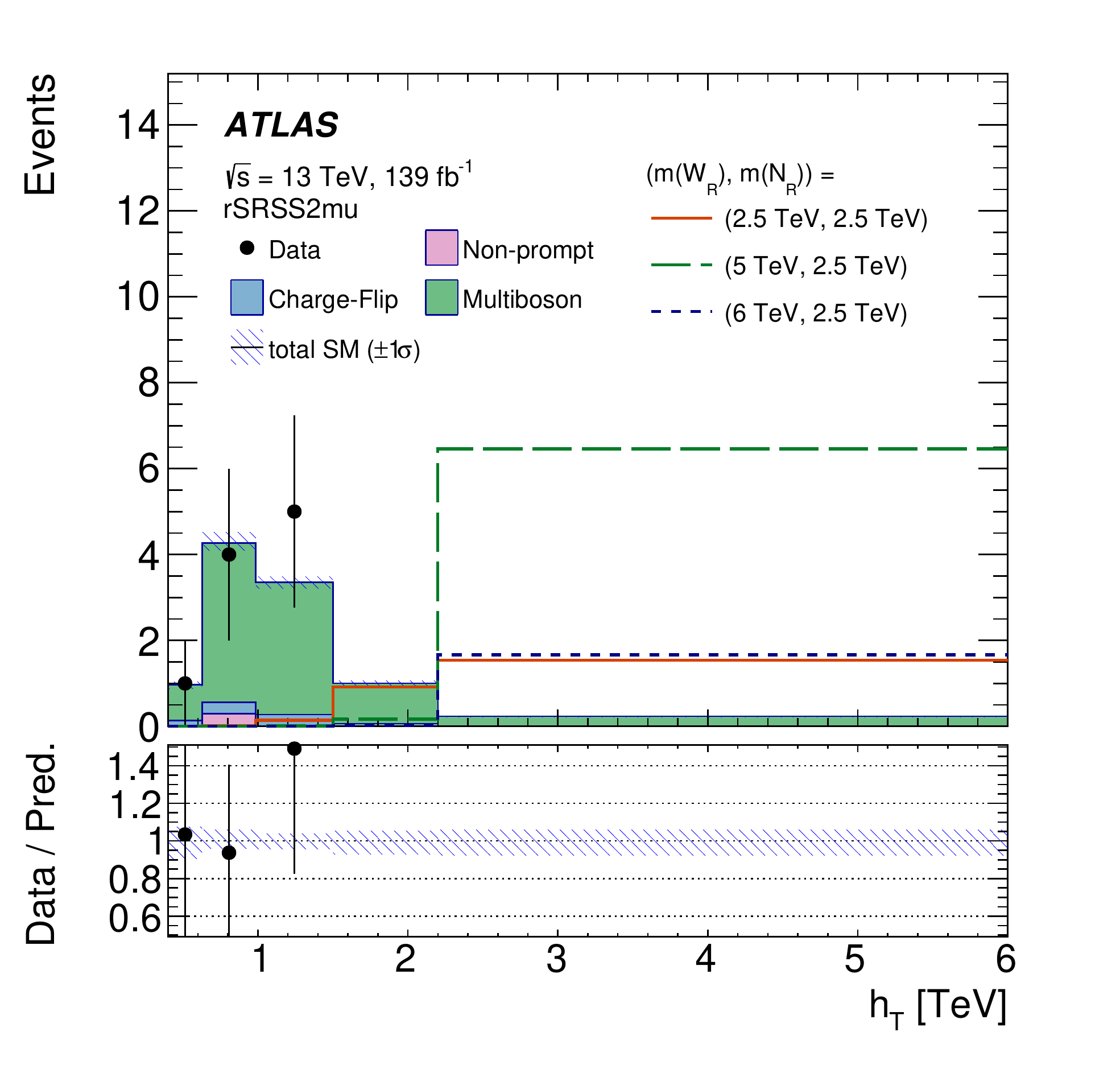}
} \\
\caption{
\texttt{SR} fits in the resolved channel: The $m_{\ell\ell jj}$ distributions in (a) \rSROSTwoe\ and (b) \rSROSTwomu, the $m_{jj}$ distributions in (c) \rSROSTwoe\ and (d) \rSROSTwomu, and the $h_{\textrm{T}}$ distributions in (e) \rSRSSTwoe\ and (f) \rSRSSTwomu.
`Top' refers to all processes containing at least one top quark. `Multiboson' refers to $VV$ and $VVV$ processes, where $V$ = $W$ or $Z$.
`Non-prompt' and `Charge-Flip' refer to processes containing a mis-identified electron and $Z$+jets with charge mis-identification.
The background expectation is the result of the fit to the \texttt{CR} data. The hatched bands (`total SM') include all systematic uncertainties post-fit with the correlation
between various sources taken into account. A blue arrow indicates an out-of-range data point for a given bin.
}
\label{fig:SRdataMC_resolved}
\end{figure}

\subsection{Boosted channel}
\label{results:boosted}
In the boosted channel, the reconstructed \mWR\ distribution ($m_{eJ}$ or $m_{\ell\ell J}$) is used exclusively since $\mWR \gg \mNR$. In the electron channel, the normalisations of $W$+jets, $Z$+jets, $\gamma$+jets and multijet samples are simultaneously estimated by the fit in the \texttt{CR}s and \texttt{SR}s. In the muon channel, only the $Z$+jet contribution is estimated by the fit.
 
A combined ``background-only'' fit is performed separately in the electron and muon channels in the \bCRWOnee, \bCRFakeOnee, \bCRyOnee, \bCRLowOnee\ and \bCRZTwoe, and \bCRZTwomu\ regions, respectively.
The normalisation factors for $W$+jets, $Z$+jets, $\gamma$+jets and multijet are free to float in the electron channel, while
only the $Z$+jets normalisation is treated as a free parameter in the muon channel.
The validity of the method is confirmed by applying the normalisation factors obtained with the fit in the \bVROnee\ and \bVRTwoe\ regions in the electron channel, and the \bVRTwomu\ region in the muon channel.
As shown in Figure~\ref{fig:boosted_mass}, a good agreement between data and the estimated SM background is found in all \texttt{CR}s and \texttt{VR}s.
 
The post-fit \mWR\ distributions for 1e, 2e and 2mu categories are found in Figure~\ref{fig:boosted_mass}.
The first bin in the 1e category shows the sum of \bCRLowOnee\ and \bCRyOnee.
Again, no significant deviation above the expected background is observed in the data.
\begin{figure}[htb]
\centering
\subfloat[]{
\includegraphics[width=0.4\textwidth, keepaspectratio]{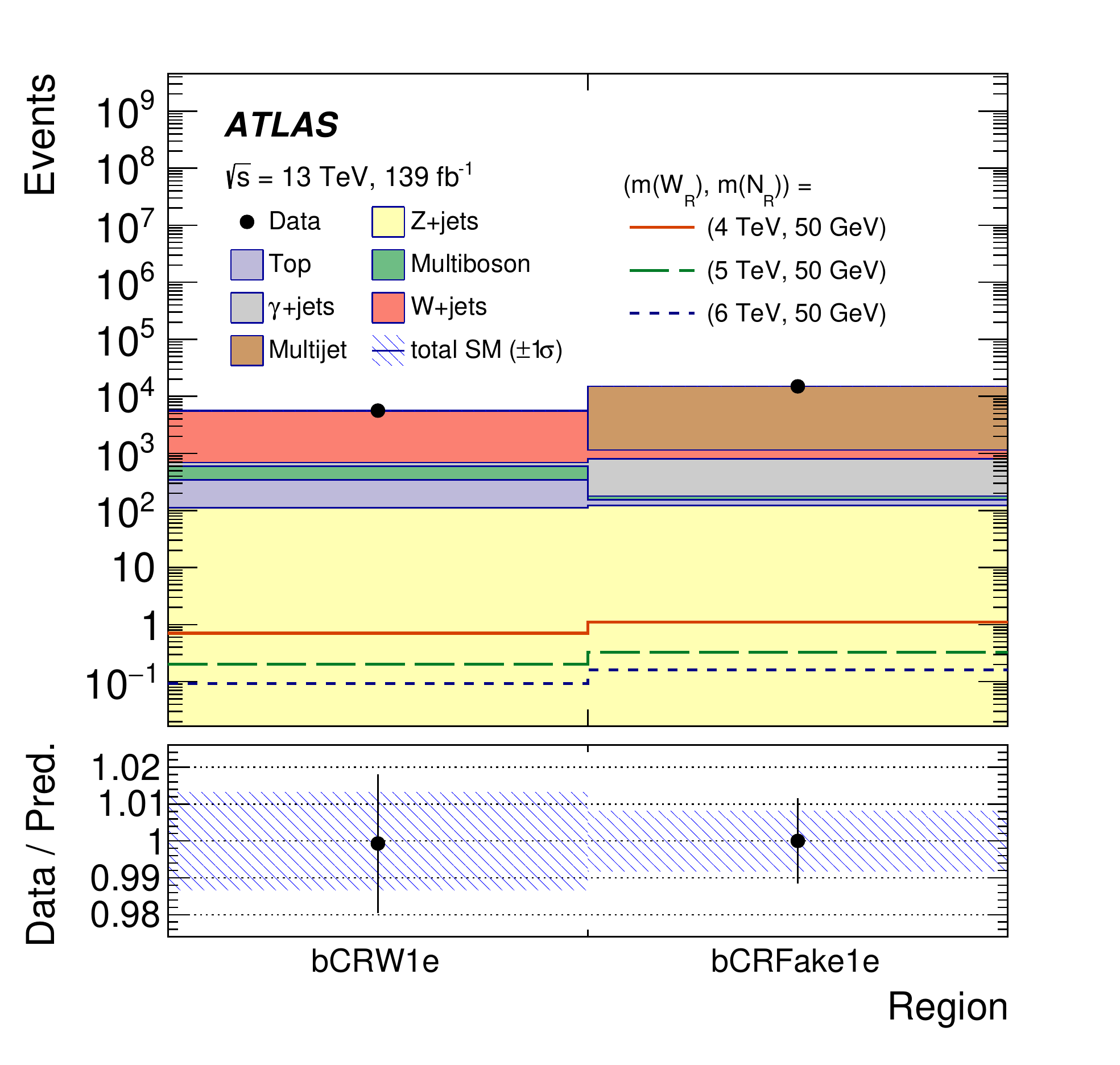}
}
\subfloat[]{
\includegraphics[width=0.4\textwidth, keepaspectratio]{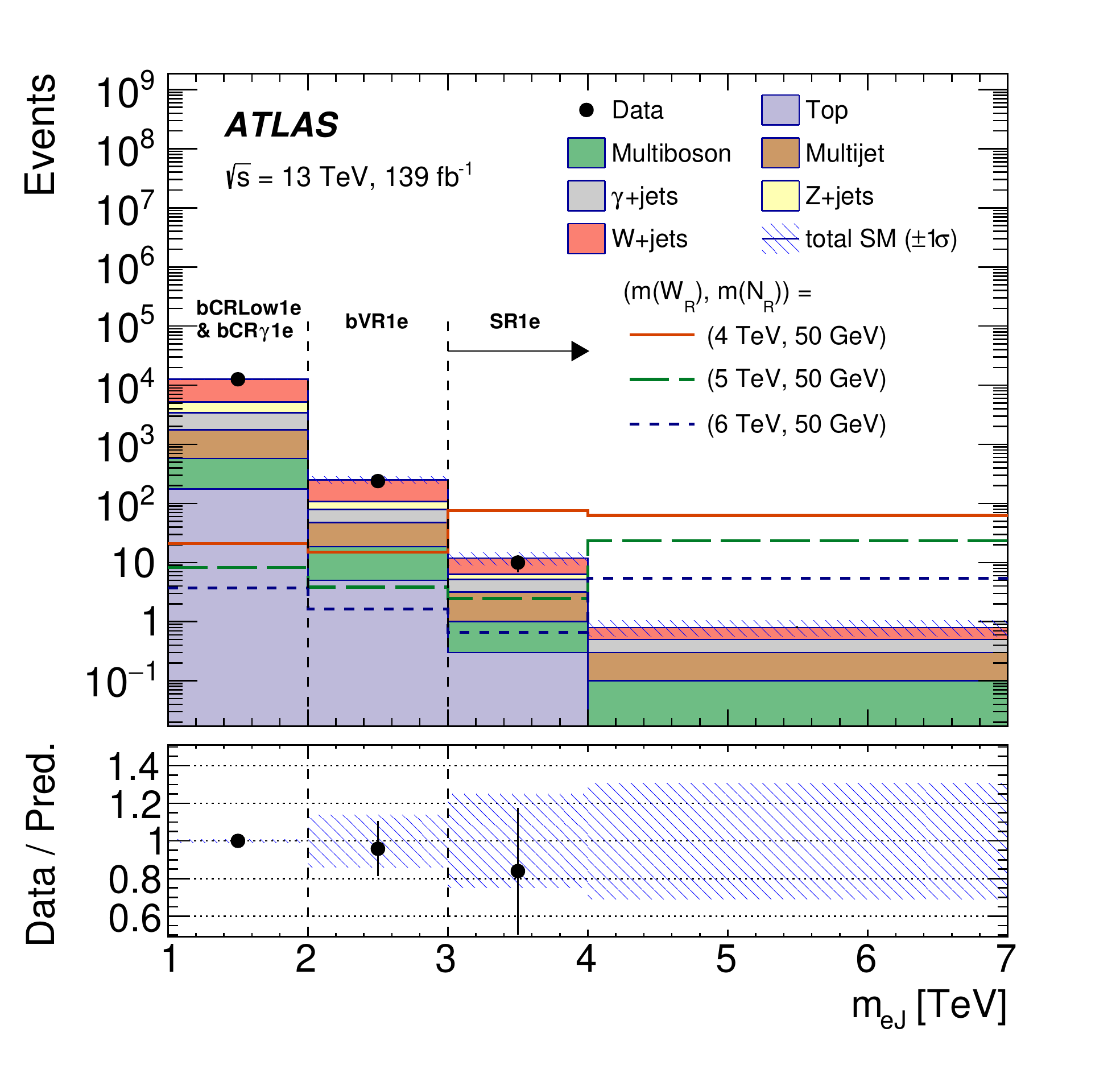}
} \\
\subfloat[]{
\includegraphics[width=0.4\textwidth, keepaspectratio]{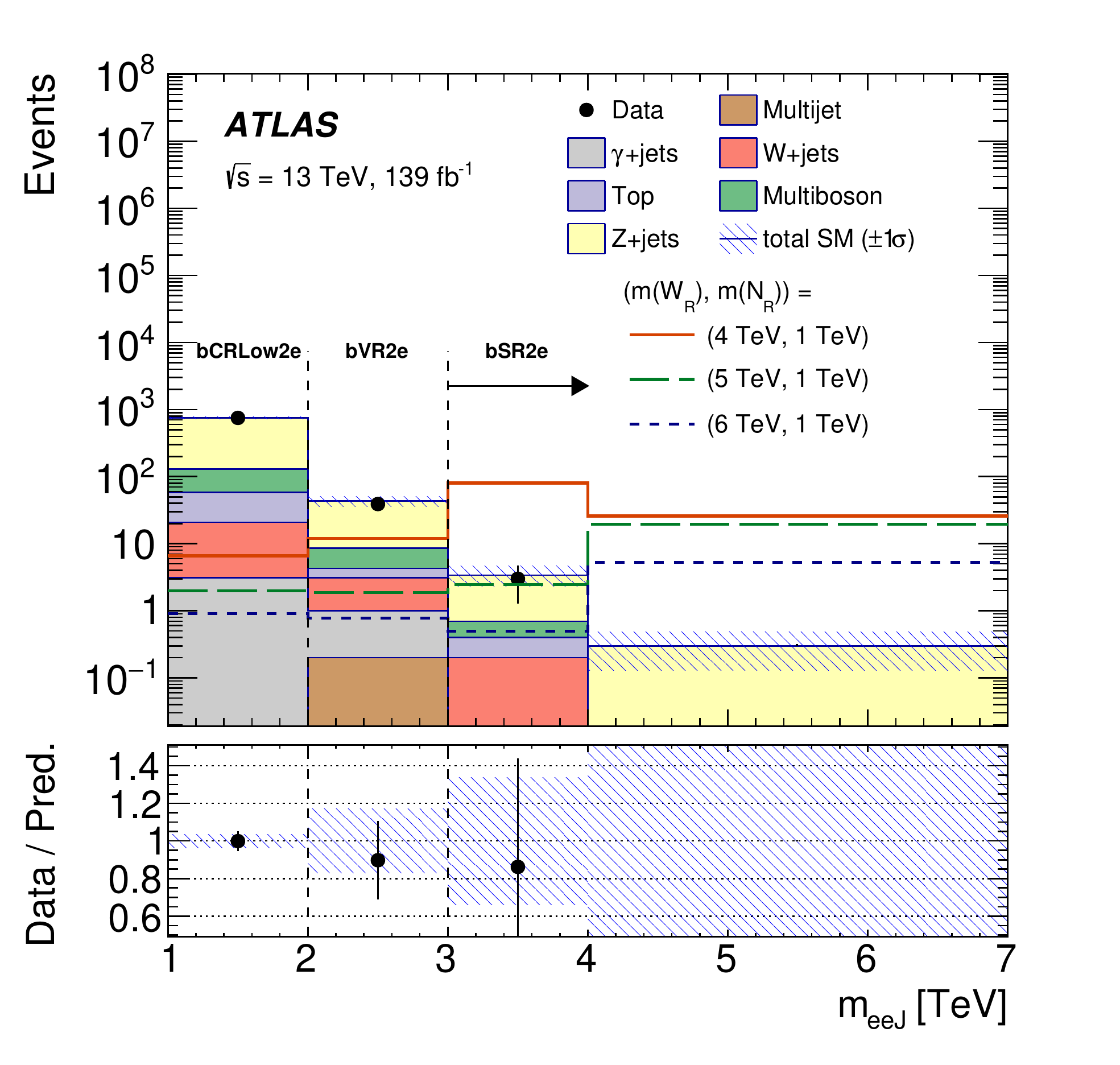}
}
\subfloat[]{
\includegraphics[width=0.4\textwidth, keepaspectratio]{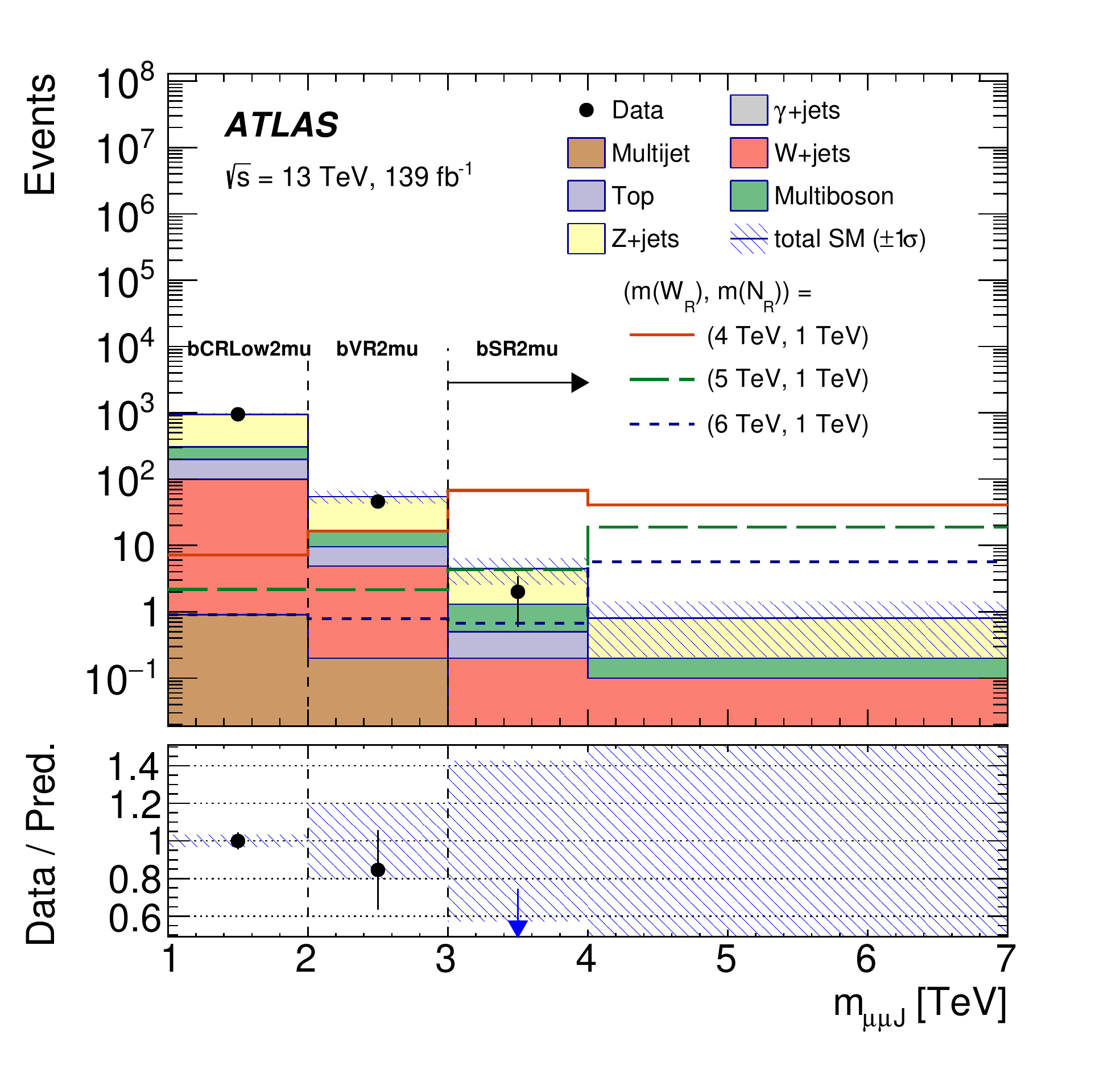}
} \\
\caption{\texttt{CR}/\texttt{VR}/\texttt{SR} fits in the boosted channel: Integrated number of events for observed data and expected background in \bCRWOnee\ and \bCRFakeOnee\ (a),
and observed \mWR\ distributions in 1e (b), 2e (c) and 2mu (d) categories. For (b)--(d), the first and second bins are used as \texttt{CR} and \texttt{VR}, and the third and fourth
bins correspond to the \texttt{SR}.
`Top' refers to all processes containing at least one top quark. `Multiboson' refers to $VV$ and $VVV$ processes, where $V$ = $W$ or $Z$.
The expected background is determined via a fit on the \texttt{CR} data.
The hatched bands (`total SM') include all post-fit systematic uncertainties, having taken into account all correlations among various sources. A blue arrow indicates an out-of-range data point for a given bin.
}
\label{fig:boosted_mass}
\end{figure}
 
Exclusion limits at 95\% CL on the signal strength of the KS process as a function of \mWR\ and \mNR\ are calculated again, with the results shown in Figure~\ref{fig:limit}. For Majorana neutrinos,
the most stringent lower limit on the \WR\ mass is 6.4~\TeV, observed in both electron and muon channels at $\mNR<1~\TeV$.
This result is an improvement over previous ATLAS searches, and extends the exclusion limits on $\mWR$ by about 1.5~TeV.
There is a particular improvement in the $\mNR < 100~\GeV$ region via the incorporation of the new \bSROnee\ region and the optimisation of \bSRTwomu.
The observed limit on $\mWR$ is  6.2 (6.1) TeV in the electron (muon) channel at $\mNR = 50~\GeV$.

\begin{figure}[phtb]
\centering
\subfloat[]{
\includegraphics[width=0.48\textwidth, keepaspectratio]{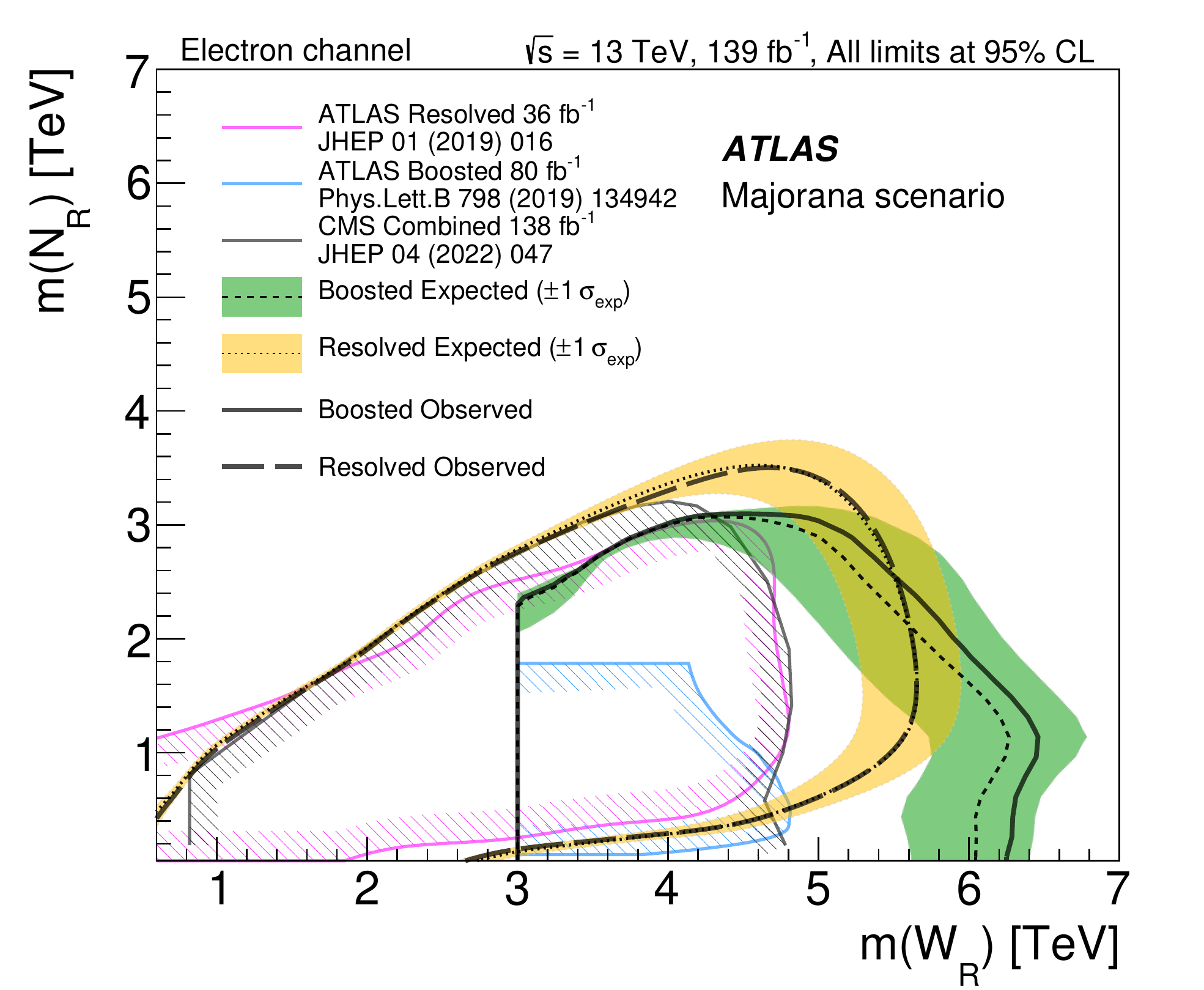}
}
\subfloat[]{
\includegraphics[width=0.48\textwidth, keepaspectratio]{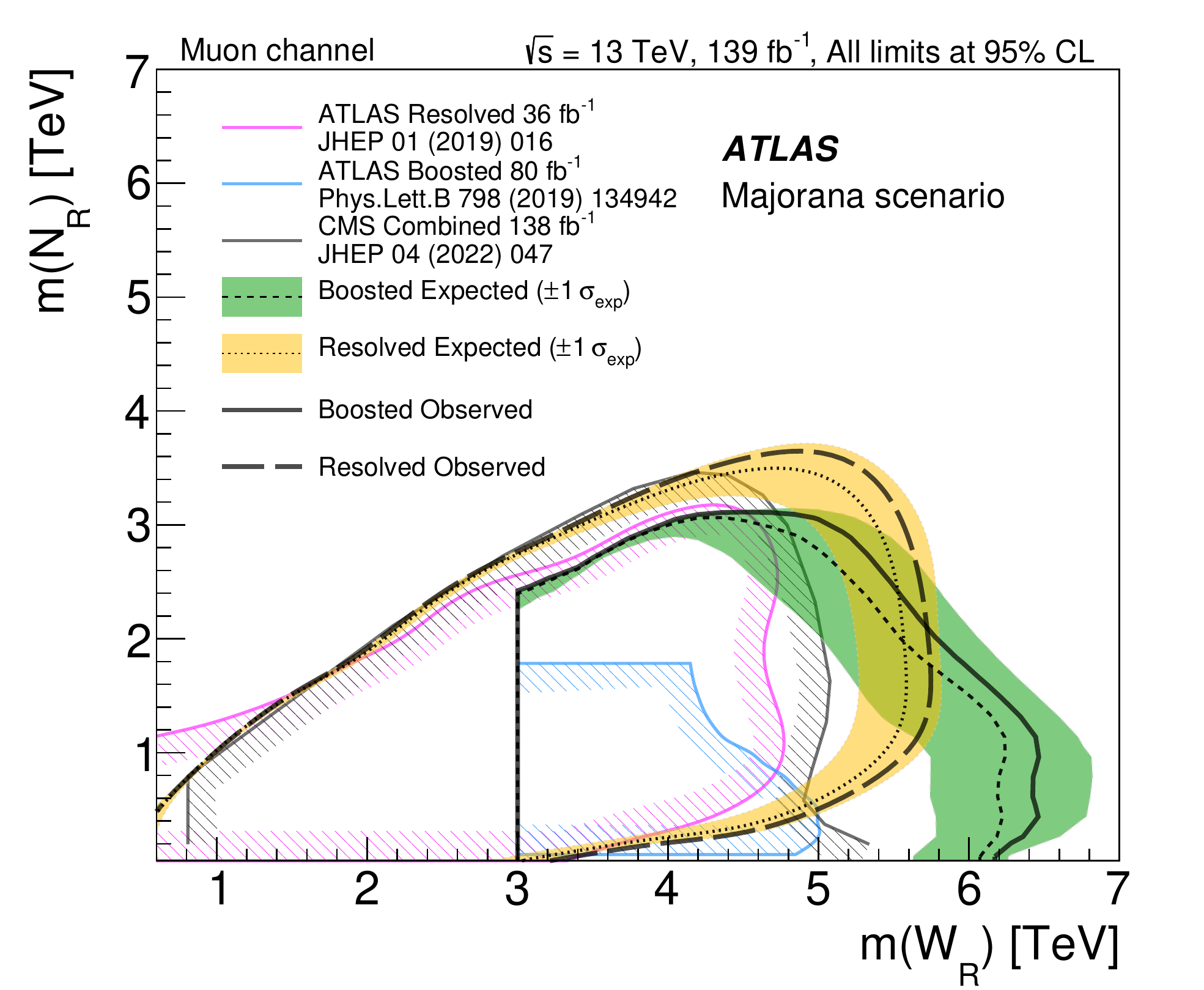}
} \\
\subfloat[]{
\includegraphics[width=0.48\textwidth, keepaspectratio]{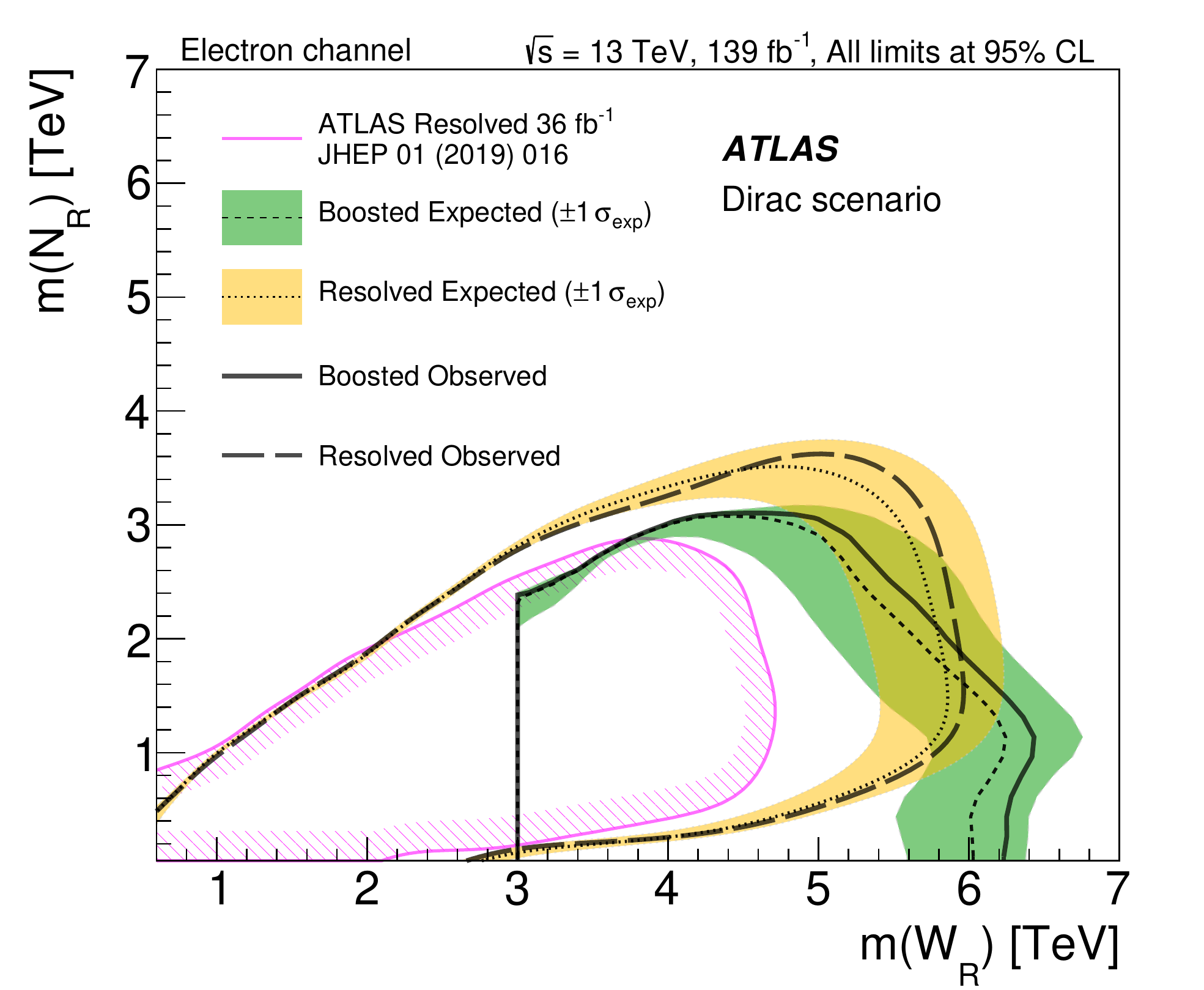}
}
\subfloat[]{
\includegraphics[width=0.48\textwidth, keepaspectratio]{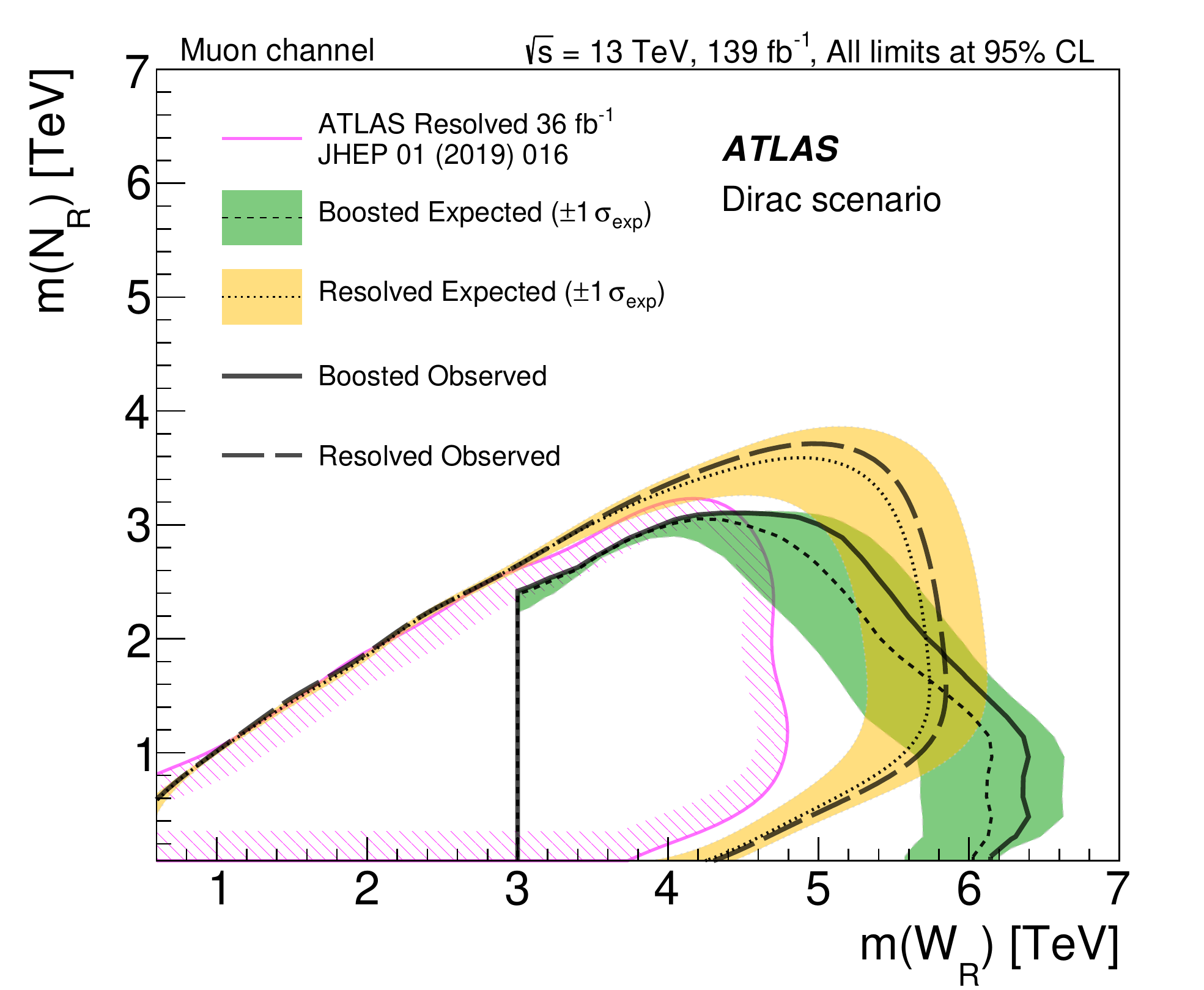}
} \\
\caption{Expected and observed 95\% CL upper limits for the Majorana (top) and Dirac (bottom) neutrino interpretations
in the electron (left) and muon (right) channels. Exclusion limits from previous ATLAS ~\cite{EXOT-2017-18,EXOT-2018-15} and CMS~\cite{CMS-EXO-20-002} searches are overlaid for comparison.
}
\label{fig:limit}
\end{figure}


\FloatBarrier
 
\section{Conclusion}

A search for right-handed \WR\ bosons and heavy right-handed Majorana or Dirac neutrinos \NR\ is presented.
The analysis uses various final states depending on the mass difference between \WR\ and \NR:
two jets and a pair of charged leptons ($\ell \ell jj$), one large-$R$ jet and two charged leptons ($\ell \ell J$), or one large-$R$ jet and one electron ($eJ$), with $\ell = e, \mu$.
It is performed with a 139~\ifb\ sample of $pp$ collisions at $\sqrt{s}$ = 13~TeV recorded by the ATLAS detector at LHC.
No evidence of \WR\ bosons or Majorana or Dirac heavy neutrinos is found assuming the KS process, and lower limits are set on $\mWR$ and $\mNR$, assuming equality of left- and right-handed weak couplings ($g_L = g_R$). The excluded region for the Majorana neutrinos extends to about $\mWR$ = 6.4~TeV for $\mNR$ =  1~TeV in both electron and muon channels. The $\mNR$ limits reach about 3.5~TeV in the electron channel and 3.6~TeV in the muon channel (for $\mWR = 4.8~\TeV$).
For Dirac neutrinos, limits reach about $\mWR$ = 6.4~TeV for $\mNR$ = 1~TeV in both electron  and muon channels.
Limits of $\mNR$ up to  3.6~TeV in the electron channel and 3.8~TeV in the muon channel for $\mWR$ = 5.0~TeV are set. The new results constitute the most stringent limits to date for the KS process.


\section*{Acknowledgements}


We thank CERN for the very successful operation of the LHC, as well as the
support staff from our institutions without whom ATLAS could not be
operated efficiently.
 
We acknowledge the support of
ANPCyT, Argentina;
YerPhI, Armenia;
ARC, Australia;
BMWFW and FWF, Austria;
ANAS, Azerbaijan;
CNPq and FAPESP, Brazil;
NSERC, NRC and CFI, Canada;
CERN;
ANID, Chile;
CAS, MOST and NSFC, China;
Minciencias, Colombia;
MEYS CR, Czech Republic;
DNRF and DNSRC, Denmark;
IN2P3-CNRS and CEA-DRF/IRFU, France;
SRNSFG, Georgia;
BMBF, HGF and MPG, Germany;
GSRI, Greece;
RGC and Hong Kong SAR, China;
ISF and Benoziyo Center, Israel;
INFN, Italy;
MEXT and JSPS, Japan;
CNRST, Morocco;
NWO, Netherlands;
RCN, Norway;
MEiN, Poland;
FCT, Portugal;
MNE/IFA, Romania;
MESTD, Serbia;
MSSR, Slovakia;
ARRS and MIZ\v{S}, Slovenia;
DSI/NRF, South Africa;
MICINN, Spain;
SRC and Wallenberg Foundation, Sweden;
SERI, SNSF and Cantons of Bern and Geneva, Switzerland;
MOST, Taipei;
TENMAK, T\"urkiye;
STFC, United Kingdom;
DOE and NSF, United States of America.
In addition, individual groups and members have received support from
BCKDF, CANARIE, CRC and DRAC, Canada;
PRIMUS 21/SCI/017 and UNCE SCI/013, Czech Republic;
COST, ERC, ERDF, Horizon 2020, ICSC-NextGenerationEU and Marie Sk{\l}odowska-Curie Actions, European Union;
Investissements d'Avenir Labex, Investissements d'Avenir Idex and ANR, France;
DFG and AvH Foundation, Germany;
Herakleitos, Thales and Aristeia programmes co-financed by EU-ESF and the Greek NSRF, Greece;
BSF-NSF and MINERVA, Israel;
Norwegian Financial Mechanism 2014-2021, Norway;
NCN and NAWA, Poland;
La Caixa Banking Foundation, CERCA Programme Generalitat de Catalunya and PROMETEO and GenT Programmes Generalitat Valenciana, Spain;
G\"{o}ran Gustafssons Stiftelse, Sweden;
The Royal Society and Leverhulme Trust, United Kingdom.
 
The crucial computing support from all WLCG partners is acknowledged gratefully, in particular from CERN, the ATLAS Tier-1 facilities at TRIUMF/SFU (Canada), NDGF (Denmark, Norway, Sweden), CC-IN2P3 (France), KIT/GridKA (Germany), INFN-CNAF (Italy), NL-T1 (Netherlands), PIC (Spain), RAL (UK) and BNL (USA), the Tier-2 facilities worldwide and large non-WLCG resource providers. Major contributors of computing resources are listed in Ref.~\cite{ATL-SOFT-PUB-2023-001}.


\printbibliography
 
\clearpage
\input{atlas_authlist}

\end{document}

%% file: atlas_authlist.tex
 
\begin{flushleft}
\hypersetup{urlcolor=black}
{\Large The ATLAS Collaboration}

\bigskip

\AtlasOrcid[0000-0002-6665-4934]{G.~Aad}$^\textrm{\scriptsize 102}$,
\AtlasOrcid[0000-0002-5888-2734]{B.~Abbott}$^\textrm{\scriptsize 120}$,
\AtlasOrcid[0000-0002-1002-1652]{K.~Abeling}$^\textrm{\scriptsize 55}$,
\AtlasOrcid[0000-0001-5763-2760]{N.J.~Abicht}$^\textrm{\scriptsize 49}$,
\AtlasOrcid[0000-0002-8496-9294]{S.H.~Abidi}$^\textrm{\scriptsize 29}$,
\AtlasOrcid[0000-0002-9987-2292]{A.~Aboulhorma}$^\textrm{\scriptsize 35e}$,
\AtlasOrcid[0000-0001-5329-6640]{H.~Abramowicz}$^\textrm{\scriptsize 151}$,
\AtlasOrcid[0000-0002-1599-2896]{H.~Abreu}$^\textrm{\scriptsize 150}$,
\AtlasOrcid[0000-0003-0403-3697]{Y.~Abulaiti}$^\textrm{\scriptsize 117}$,
\AtlasOrcid[0000-0003-0762-7204]{A.C.~Abusleme~Hoffman}$^\textrm{\scriptsize 137a}$
% ...
\bigskip
\\

% ...
$^{29}$Physics Department, Brookhaven National Laboratory, Upton NY; United States of America.\\
% ...
$^{35}$$^{(a)}$Facult\'e des Sciences Ain Chock, R\'eseau Universitaire de Physique des Hautes Energies - Universit\'e Hassan II, Casablanca;$^{(b)}$Facult\'{e} des Sciences, Universit\'{e} Ibn-Tofail, K\'{e}nitra;$^{(c)}$Facult\'e des Sciences Semlalia, Universit\'e Cadi Ayyad, LPHEA-Marrakech;$^{(d)}$LPMR, Facult\'e des Sciences, Universit\'e Mohamed Premier, Oujda;$^{(e)}$Facult\'e des sciences, Universit\'e Mohammed V, Rabat;$^{(f)}$Institute of Applied Physics, Mohammed VI Polytechnic University, Ben Guerir; Morocco.\\
% ...
$^{49}$Fakult\"{a}t Physik , Technische Universit{\"a}t Dortmund, Dortmund; Germany.\\
% ...
$^{55}$II. Physikalisches Institut, Georg-August-Universit\"{a}t G\"ottingen, G\"ottingen; Germany.\\
% ...
$^{102}$CPPM, Aix-Marseille Universit\'e, CNRS/IN2P3, Marseille; France.\\
% ...
$^{117}$Department of Physics, New York University, New York NY; United States of America.\\
% ...
$^{120}$Homer L. Dodge Department of Physics and Astronomy, University of Oklahoma, Norman OK; United States of America.\\
% ...
$^{137}$$^{(a)}$Departamento de F\'isica, Pontificia Universidad Cat\'olica de Chile, Santiago;$^{(b)}$Millennium Institute for Subatomic physics at high energy frontier (SAPHIR), Santiago;$^{(c)}$Instituto de Investigaci\'on Multidisciplinario en Ciencia y Tecnolog\'ia, y Departamento de F\'isica, Universidad de La Serena;$^{(d)}$Universidad Andres Bello, Department of Physics, Santiago;$^{(e)}$Instituto de Alta Investigaci\'on, Universidad de Tarapac\'a, Arica;$^{(f)}$Departamento de F\'isica, Universidad T\'ecnica Federico Santa Mar\'ia, Valpara\'iso; Chile.\\
% ...
$^{150}$Department of Physics, Technion, Israel Institute of Technology, Haifa; Israel.\\
$^{151}$Raymond and Beverly Sackler School of Physics and Astronomy, Tel Aviv University, Tel Aviv; Israel.\\
% ...
$^{a}$ Also Affiliated with an institute covered by a cooperation agreement with CERN.\\
$^{b}$ Also at An-Najah National University, Nablus; Palestine.\\
$^{c}$ Also at Borough of Manhattan Community College, City University of New York, New York NY; United States of America.\\
$^{d}$ Also at Center for High Energy Physics, Peking University; China.\\
$^{e}$ Also at Center for Interdisciplinary Research and Innovation (CIRI-AUTH), Thessaloniki; Greece.\\
$^{f}$ Also at Centro Studi e Ricerche Enrico Fermi; Italy.\\
% ...

\end{flushleft}
